\documentclass[acmsmall,screen,nonacm]{acmart}
\AtBeginDocument{%
  }
\setcopyright{none}
\usepackage{booktabs,xspace}   
\usepackage{subcaption} 
\usepackage{multirow, rotating, amsmath, wasysym}
\usepackage[ruled,linesnumbered,vlined]{algorithm2e}
\usepackage{hyperref}
\usepackage{soul}

\mathchardef\Gamma="0100 \mathchardef\Delta="0101
\mathchardef\Theta="0102 \mathchardef\Lambda="0103
\mathchardef\Xi="0104 \mathchardef\Pi="0105
\mathchardef\Sigma="0106 \mathchardef\Upsilon="0107
\mathchardef\Phi="0108 \mathchardef\Psi="0109
\mathchardef\Omega="010A

\newcommand{\outline}[1]{}

\usepackage{algpseudocode}
\usepackage{xspace}
\usepackage{url}
\usepackage{graphicx}
\usepackage{latexsym}
\usepackage{amsfonts}
\usepackage{psfrag}
\usepackage{wrapfig}
\usepackage{comment}
\usepackage{subcaption}
\usepackage{alltt}
\usepackage{color}
\usepackage{tikz}

\newcommand{\ie}{i.e.\xspace}
\newcommand{\eg}{e.g.\xspace}

\newtheorem{theorem}{Theorem}[section]

\usepackage{listings}
\newcommand{\var}[1]{\mathnormal{#1}}

\newcommand{\field}[1]{\mathnormal{#1}}
\newcommand{\fieldset}{\mathbb{F}}
\newcommand{\prog}[1]{\texttt{#1}}

\newcommand{\pelement}[1]{\mathfrak{#1}}
\newcommand{\pele}{\pelement{a}}

\newcommand{\peleset}{\mathfrak{A}}

\newcommand{\critset}{\mathcal{S}}
\newcommand{\nonconst}{\pi}

\newcommand{\allvarset}{\mathbb{V}}
\newcommand{\absloc}{\omega}

\newcommand{\locationset}{\mathbb{L}}

\newcommand{\freevariables}{\mathbf{free}}

\newcommand{\gamapointer}{\gamma}

\newcommand{\penv}{\mathbf{eval}}
\newcommand{\pointsto}{\mathbf{pt}}

\newcommand{\pjoin}{\sqcup}

\newcommand{\pdispatch}{\mathbf{dispatch}}

\newcommand{\syntax}[1]{\mathtt{#1}}
\newcommand{\syntProg}{\mathtt{P}}
\newcommand{\syntStmt}{\mathtt{stmt}}
\newcommand{\syntStmtset}{\mathbb{S}}
\newcommand{\syntStmts}{\mathcal{S}}

\newcommand{\mapproctostmt}{\Gamma}
\newcommand{\syntlv}{\syntax{lv}}
\newcommand{\syntexpr}{\syntax{expr}}

\newcommand{\concSet}{\Psi}
\newcommand{\concEle}{\sigma}

\newcommand{\const}{c}
\newcommand{\constset}{\mathbb{C}}
\newcommand{\syntloc}{l}
\newcommand{\syntlocset}{\mathbb{L}}

\newcommand{\predcrit}{\Phi}

\newcommand{\ciele}{\mathfrak{a}}

\newcommand{\cieleset}{\mathfrak{A}}

\newcommand{\fixpoint}{\mathbf{lfp}}

\newcommand{\node}{\kappa}

\newcommand{\tkzputform}[2]{\node[] at (#1){\ensuremath{#2}};}

\newbox\subfigbox 
\makeatletter
{\def\caption##1{\gdef\subcapsave{\relax##1}}%
\let\subcapsave=\@empty 
\let\sf@oldlabel=\label
\def\label##1{\xdef\sublabsave{\noexpand\label{##1}}}%
\let\sublabsave\relax 
\setbox\subfigbox\hbox
\bgroup}
{\egroup 
\let\label=\sf@oldlabel
\subfigure[\subcapsave]{\box\subfigbox}}%
\makeatother

\newcommand{\cSem}[1]{\llbracket#1\rrbracket}

\newcommand{\absSemTop}[1]{\llbracket#1\rrbracket^{t}}
\newcommand{\absSemCom}[1]{\llbracket#1\rrbracket^{c}}
\newcommand{\absSemHybrid}[1]{\llbracket#1\rrbracket^{h}}

\newcommand{\absEleTop}{\mathfrak{a}}
\newcommand{\absEleCom}{\mathfrak{c}}
\newcommand{\absEleHybrid}{\mathfrak{h}}
\newcommand{\absEleTopSet}{\mathfrak{A}}
\newcommand{\absEleComSet}{\mathfrak{C}}

\newcommand{\transTop}{\mathbf{trans}^{t}}
\newcommand{\transCom}{\mathbf{trans}^{c}}
\newcommand{\compose}{\mathbf{comp}^{c}}

\newcommand{\topJoin}{\sqcup^{t}}
\newcommand{\comJoin}{\mathbf{join}^{c}}
\newcommand{\comJoinI}{\sqcup^{c}}
\newcommand{\hybridJoin}{\sqcup^{h}}
\newcommand{\sumCom}{\mathbf{sum}^{c}}
\newcommand{\sumHybrid}{\mathbf{sum}^{h}}
\newcommand{\apply}{\mathbf{apply}^{h}}
\newcommand{\predicateSen}{\mathcal{P}}

\newcommand{\valueSet}{\mathbf{val}}

\newcommand{\progToSet}{\Lambda}

\newcommand{\toolname}{Hybrid Inlining\xspace}

\newcommand{\contextReady}{\mathbf{ready}}
\newcommand{\procCall}{\syntax{call\; proc}}
\newcommand{\pCom}{\absEleCom}
\newcommand{\consGenP}{\mathbf{cons}}

\newcommand{\comJoinP}{\sqcup^{c}}
\newcommand{\pointerTrans}{\mathbf{trans}}
\usepackage{makecell}
\newcommand{\presec}{\vspace{-0.0in}}
\newcommand{\postsec}{\vspace{-0.0in}}

\usepackage{fancyhdr}
\renewcommand\footnotetextcopyrightpermission[1]{} 
\begin{document}
\title[\toolname]{\toolname: A Compositional and Context Sensitive Static Analysis Framework}

\author{Jiangchao Liu}                                  
\affiliation{
  \institution{Ant Group}         
    \city{HangZhou}
  \country{China}
}
\email{jiangchao.ljc@jantgroup.com}          

\author{Jierui Liu}                                  
\affiliation{
  \institution{Ant Group}            
      \city{HangZhou}
  \country{China}
}
\email{liujierui.ljr@antgroup.com}          

\author{Peng Di}                                  
\affiliation{
  \institution{Ant Group}            
      \city{HangZhou}
  \country{China}
}
\email{dipeng.dp@antgroup.com}          

\author{Diyu Wu}                                  
\affiliation{
  \institution{Ant Group}            
      \city{HangZhou}
  \country{China}
}
\email{wudiyu.wdy@antgroup.com}          

\author{Hengjie Zheng}                                  
\affiliation{
  \institution{Ant Group}            
      \city{HangZhou}
  \country{China}
}
\email{zhenghengjie.zhj@antgroup.com}          

\author{Alex Liu}                                  
\affiliation{
  \institution{Ant Group}            
      \city{HangZhou}
  \country{China}
}
\email{alexliu@antgroup.com}          

\author{Jingling Xue}                                  
\affiliation{
  \institution{UNSW}            
      \city{Sydney}
  \country{Australia}
}
\email{jingling@cse.unsw.edu.au}          


\begin{abstract}

Context sensitivity is essential for achieving the precision in inter-procedural static analysis. To be (fully) context sensitive, top-down analysis needs to fully inline all statements of the callees at each callsite, leading to statement explosion. Compositional analysis, which inlines summaries of the callees,  scales up but often loses precision, as it is not strictly context sensitive.
We propose a compositional and strictly context sensitive framework for static analysis. This framework is based on one key observation: a compositional static analysis often loses precision only on some \emph{critical} statements that need to be analyzed context sensitively. Our approach hybridly inlines the critical statements and the summaries of non-critical statements of each callee, thus avoiding the re-analysis of non-critical ones.
In addition, our analysis lazily summarizes the critical statements, by stopping propagating the critical statements once the calling context accumulated is adequate.
Hybrid Inlining can be as precise as context sensitive top-down analysis. We have designed and implemented a pointer analysis based on this framework. It can analyze large Java programs from the Dacapo benchmark suite and industry in minutes. In our evaluation, compared to context insensitive analysis, Hybrid Inlining just brings 65\% and 1\% additional time overhead on Dacapo and industrial applications respectively.
\end{abstract}


\maketitle
\presec \section{Introduction} \label{sec:introduction} \postsec

%
Context sensitivity is essential for the precision of inter-procedural static analysis.
There are two ways to achieve inter-procedural  analysis: top-down and compositional (i.e., bottom-up).
They both fall short with respect to context sensitivity in analyzing software at industrial scale.

To be context sensitive, a top-down analysis needs to inline explicitly or implicitly  the statements of the called procedures at each callsite.
Given a large industrial software application,  inlining all its procedures often results in statement explosion, especially when polymorphism is heavily used.
A compromise is to limit context sensitivity, such as \emph{k}-callsite sensitivity~\cite{shivers1988control} and  \emph{k}-object sensitivity~\cite{milanova2005parameterized}.
However, this results in precision loss~\cite{mangal2014correspondence}.
Scalability with high context sensitivity remains challenging even though recent studies have made significant progress with techniques like parallelization~\cite{albarghouthi2012parallelizing}, demand driven~\cite{sridharan2005demand,arzt2014flowdroid,heintze2001demand}, selective context sensitivity~\cite{jeong2017data,li2020principled,lu2019precision}, and incremental analysis~\cite{shang2012fast}.

A compositional, \ie, bottom-up approach~\cite{cousot2002modular} starts from the leaf procedures of the call graph, and inlines the summaries of the called procedures at each callsite.
Since each procedure is analyzed only once to obtain its summary, it is much easier for the
analysis to scale up.
However, the summaries are computed with all possible input states (\ie, \emph{top}) as a precondition. Inlining such summaries often leads to less precise results than inlining statements.
This makes compositional analysis non-strictly context sensitive.
%
%
Some works tackle context sensitivity with subtle designs of summaries~\cite{dillig2011precise,feng2015bottom,calcagno2009compositional,nystrom2004bottom,DSA:PLDI07}.
However, these works either are specific to a kind of properties or introduce disjunction which brings potential disjunctive explosion. 


In this paper, we propose a compositional and context sensitive static analysis framework called \emph{\toolname}. Our key observation is that the summarization process in a compositional static analysis is imprecise only on some statements due to the absence of contexts, referred to as
\emph{critical} statements.
In other words, the summarization of non-critical statements with/without contexts generates the same postcondition for any precondition.
%
%
At each callsite, our framework hybridly inlines the critical statements (for context sensitivity) and the summaries of non-critical statements (for performance) of the called procedures.
Thus, the critical statements are propagated from callees to the contexts in their callers. 
The propagation of a critical statement stops  at a callsite where the context is \emph{adequate} (since no further precision improvement with more propagation is possible). 
Then the critical statement is summarized.
Compositional analysis with \toolname can be as precise as an \(\infty\)-callsite sensitive top-down analysis.
Compared with conventional compositional context insensitive analysis, the non-critical statements are still analyzed once and for all, but the critical statements are analyzed multiple times, depending on the distinct contexts they encounter.


The key difference of our approach with all other compositional analyses~\cite{dillig2011precise,feng2015bottom,calcagno2009compositional,nystrom2004bottom,DSA:PLDI07} is that, our approach is about \emph{when} to summarize the statements in a program, rather than \emph{how} to summarize them. 

Our approach can be seen as an application of selective context sensitivity on compositional analysis. While the idea of selective context sensitivity has been well explored in top-down analysis~\cite{li2020principled,lu2019precision}, this paper is the first attempt on compositional analysis. Thanks to the fact that compositional analysis can  summarize each statement without the context, our approach enables fine-grained selective context sensitivity in the level of statements rather than the level of procedures~\cite{li2020principled,lu2019precision}.  Another advantage of our approach is that the propagation of  critical statements stops whenever the context is adequate, not necessarily at the root procedures. This avoids unnecessary propagation without sacrificing the ability to allow individual critical statements to be propagated far.





%
%
We have designed and implemented a non-trivial compositional pointer analysis for Java programs.
While compositional analysis is known for easy scalability, \toolname enhances it with context sensitivity at almost no cost on performance. 
In practice, the compositional and context sensitive pointer analysis can analyze large  programs from Dacapo~\cite{blackburn2006dacapo} and industry in minutes on a personal laptop.  
%


In this paper, we make  the following three key contributions:
\begin{itemize}
\item We propose a \toolname framework for performing compositional analysis, which brings context sensitivity at little cost on performance.

\item We instantiate it with a non-trivial pointer analysis.

\item We share our experience in applying the pointer analyses to  10 open source benchmarks from Dacapo and 10 real industrial large benchmarks. With \toolname, the context sensitive analysis runs on large programs as fast as context insensitive analysis.



\end{itemize}

\section{Overview} \label{sec:overview}

In this section, we illustrate the key idea of our \toolname framework by a pointer analysis on the Java code in Figure~\ref{fig:motivating}. With this example, we first consider  a conventional compositional flow insensitive pointer analysis, which is context insensitive. Then we demonstrate  how \toolname improves its precision with strict context sensitivity. 

In this example, an interface \prog{X} (line 2) declares an abstract procedure \prog{poly}(). Two classes \prog{Y} and \prog{Z} both implement \prog{X} and override the abstract procedure \prog{poly}() differently. For clarity, we use  \prog{polyY}() and  \prog{polyZ}() to denote the procedure \prog{poly}() in \prog{Y} and \prog{Z} respectively, where \prog{polyY} is an identity procedure and \(\prog{polyZ}\) just returns a new instance of the class \prog{Obj}. In the class \prog{FacadeImpl}, an identity procedure \prog{id}() at line 18 is defined. The procedure \prog{foo}() calls the abstract procedure \prog{poly}() (at line 25), the implementation of which is only known when its first parameter is decided. The procedures  \prog{bar}1() and \prog{bar}2() call \prog{foo}() indirectly with the receiver of the procedure \prog{poly}() specialized as instances of \prog{Y} and \prog{Z}, respectively. Despite the  complexity in their code, \prog{bar}1() can actually be seen as an identity procedure and \prog{bar}2() returns a new instance of the class \prog{Obj}. The procedure \prog{service}() is the root procedure. It is obvious that the two assertions in it are always \(true\). To verify them with static analysis, context sensitivity is needed. 

\begin{center}

\newsavebox{\overcodeone} 
\begin{lrbox}{\overcodeone}
\begin{minipage}{.45\textwidth}
\footnotesize
\begin{verbatim}
1  public class Obj{}
2  public interface X{
3   Obj poly(Obj obj);
4  }
5  public class Y implements X{
6   @Override
7   public Obj poly(Obj obj){
8    return obj;
9   }
10 }
11 public class Z implements X{
12  @Override
13  public Obj poly(Obj obj){
14    return new Obj();
15  }
16 }
17 public class FacadeImpl{
18  public X id(X x){
19   X tv = x;
20   return tv;
21  }
\end{verbatim}
\end{minipage}
\end{lrbox}%

\newsavebox{\overcodetwo} 
\begin{lrbox}{\overcodetwo}
\begin{minipage}{.45\textwidth}
\footnotesize
\begin{verbatim}
22  public Obj foo(X x, Obj obj){
23    ...
24    X tx = id(x);
25    return tx.poly(obj);   
26  }
27  public Obj mid(X x, Obj obj){
28    return foo(x, obj);
29  }
30  public Obj bar1(Obj obj){
31    return mid(new Y(), obj);
32  }
33  public Obj bar2(Obj obj){
34    return mid(new Z(), obj);
35  }
36  public void service(){
37    Obj first = new Obj();
38    Obj second = bar1(first);
39    Obj third = bar2(first);
40    assert(first == second);
41    assert(first != third);
42  }
43 }
\end{verbatim}
\end{minipage}
\end{lrbox}%

\begin{figure}
    \centering
\subfloat{\raisebox{-1cm}{\usebox{\overcodeone}}}
\subfloat{\usebox{\overcodetwo}}
    \caption{An illustrating example.}
    \label{fig:motivating}
\end{figure}
    
\end{center}

\subsection{A Compositional and  Context Insensitive  Pointer Analysis}
\label{sec:overviewci}

\emph{Abstract states}. An abstract state is a computer representable abstraction of a set of concrete program states. In this pointer analysis, 
we define an abstract state \(\absEleTop \in\absEleTopSet\) as a set of set constraints on program variables (denoted as \(v\in \allvarset\)) and locations (i.e., allocation sites, denoted as  \(l\in\locationset\)) of the form \( v \supseteq v' \) and \( v \supseteq \{l\}\). For instance, \(tv\supseteq x\) (from line 19) means  that the set of locations pointed to by \(\var{tv}\) is a superset of those pointed to by \(\var{x}\). The constraint \(first \supseteq \{l_{37}\}\) (from line 37) indicates that the variable \prog{first} may point to the allocation site declared at line 37. By convention, there is a special abstract state \(\top\) that represents all possible concrete states.

\emph{Abstract summaries}. A compositional analysis is built on abstract summaries \(\absEleCom\in \absEleTopSet \times \absEleTopSet\), which are relations on abstract states. It maps a given precondition to a postcondition. A compositional analysis can summarize any program into an abstract summary,  achieved with summarization. 

\begin{figure}[ht]
  \centering
  \begin{tabular}{  c  c c c}
    	\raisebox{0.1\height}{ \scalebox{1}{  \begin{tikzpicture}[
roundnode/.style={circle, draw=green!60, fill=green!5},
squarednode/.style={rectangle, draw=black, fill=yellow!5},
]
  \begin{scope}[shift={(0, 0)}]

\node[squarednode, rounded corners]   at  (1,3)    (par)                              {\(\var{par}_{1}\)};
\node[squarednode, rounded corners]   at  (0,3)    (par0)                              {\(\var{par}_{0}\)};
\node[squarednode, rounded corners]   at  (0,2)    (this)                              {\(\var{this}\)};
\node[squarednode,rounded corners]     at  (1,2)      (obj)                            {\(\var{x}\)};
\node[squarednode,rounded corners]   at  (1,1)      (to)                            {\(\var{tv}\)};
\node[squarednode,rounded corners]     at  (1,0)      (ret)                            {\(\var{ret}\)};
\draw[thick, -latex] (par.south) -- (obj.north);
\draw[thick, -latex] (par0.south) -- (this.north);
\draw[thick, -latex] (obj.south) -- (to.north);
\draw[thick, -latex] (to.south) -- (ret.north);
  \end{scope}
\end{tikzpicture}}}
    & 
    \raisebox{.7\height}{ \scalebox{1}{  \begin{tikzpicture}[
roundnode/.style={circle, draw=green!60, fill=green!5},
squarednode/.style={rectangle, draw=black, fill=yellow!5},
]
  \begin{scope}[shift={(0, 0)}]

\node[squarednode, rounded corners]   at  (1,1)    (par)                              {\(\var{par}_{1}\)};
\node[squarednode, rounded corners]   at  (0,1)    (par0)                              {\(\var{par}_{0}\)};
\node[squarednode,rounded corners]     at  (1,0)      (ret)                            {\(\var{ret}\)};
\draw[thick, -latex] (par.south) -- (ret.north);
  \end{scope}
\end{tikzpicture}}}
    &
          \raisebox{0.7\height}{ \scalebox{1}{  \begin{tikzpicture}[
roundnode/.style={circle, draw=green!60, fill=green!5},
squarednode/.style={rectangle, draw=black, fill=blue!5},
]
  \begin{scope}[shift={(0, 0)}]

\node[squarednode, rounded corners ]   at  (1,1)    (par)                              {\(\var{par}_{1}\)};
\node[squarednode, rounded corners]     at  (1,0)      (ret)                            {\(\var{ret}\)};
\draw[-latex,thick] (par.south) -- (ret.north);
  \end{scope}
\end{tikzpicture}}}
    &
    \multirow{7}{*}[1.1in]{ 
     \raisebox{0.7\height}{ \scalebox{1}{  \begin{tikzpicture}[
roundnode/.style={circle, draw=black, fill=green!5},
squarednode/.style={rectangle, draw=black, fill=red!5},
squaredid/.style={rectangle, draw=black, fill=yellow!5},
squaredpolyy/.style={rectangle, draw=black, fill=blue!5},
squaredpolyz/.style={rectangle, draw=black, fill=green!5},
]
  \begin{scope}[shift={(0, 0)}]

\node[squarednode,rounded corners]   at  (1,3)    (parfoo2)                              {\(\var{par}_{2}\)};
\node[squarednode,rounded corners]   at  (0,3)    (parfoo1)                              {\(\var{par}_{1}\)};
\node[squarednode,rounded corners]   at  (-1,3)    (parfoo0)                              {\(\var{par}_{0}\)};
\node[squarednode,rounded corners]   at  (1,1)    (obj)                              {\(\var{obj}\)};
\node[squarednode,rounded corners]   at  (0,2)    (x)                              {\(\var{x}\)};
\node[squarednode,rounded corners]   at  (-1,2)    (this)                              {\(\var{this}\)};
\node[squaredid,rounded corners]   at  (0,1)    (parid1)                      {\(\var{par}_{1}\)};
\node[squaredid,rounded corners]   at  (-1,1)    (parid0)                      {\(\var{par}_{0}\)};
\node[squaredid,rounded corners]   at  (0,0)    (retid)                              {\(\var{ret}\)};
\tkzputform{-1,0}{\var{@id}}
 \draw[rounded corners,dashed] (-1.5,-0.5) rectangle (0.5,1.5);
\node[squarednode,rounded corners]   at  (0,-1)    (tx)                              {\(\var{tx}\)};
\node[squaredpolyy,rounded corners]   at  (1,-2)    (parY1)                              {\(\var{par}_1\)};
\tkzputform{0.4,-2.6}{@\var{polyY}}
 \draw[rounded corners,dashed] (-1.5,-4) rectangle (-0.3,-1.5);

 \draw[rounded corners,dashed] (-0.2,-3.5) rectangle (1.58,-1.5);
\tkzputform{-0.9,-3.8}{@\var{polyZ}}
\node[roundnode,rounded corners] at (-1,-2) (loc) {\(l_{14}\)};
\node[squaredpolyy,rounded corners]   at  (1,-3)    (retY)                              {\(\var{ret}\)};
\node[squaredpolyz,rounded corners]   at  (-1,-3)    (retZ)                              {\(\var{ret}\)};
\node[squarednode,rounded corners]   at  (1,-4)    (ret)                              {\(\var{ret}\)};

\draw[thick,-latex] (parfoo2.south) -- (obj.north);
\draw[thick,-latex] (parfoo1.south) -- (x.north);
\draw[thick,-latex] (x.south) -- (parid1.north);
\draw[thick,-latex] (parfoo0.south) -- (this.north);
\draw[thick,-latex] (this.south) -- (parid0.north);
\draw[thick,-latex] (parid1.south) --(retid.north);
\draw[thick,-latex] (retid.south) -- (tx.north);
 \draw[thick,-latex] (obj.south) -- (parY1.north);
 \draw[thick,-latex] (parY1.south) -- (retY.north);
 \draw[thick,-latex] (loc.south) -- (retZ.north);
 \draw[thick,-latex] (retY.south) -- (ret.north);
 \draw[thick,-latex] (retZ.south) -- (ret.north);

  \end{scope}
  \end{tikzpicture}}}
    }
    \\
    (a) \prog{id}()
    &
    (b) \prog{id}()
    &
    (c) \prog{polyY}()
    &
    \\
    &&
    \\
                     \\
        &
        &&
        \\
                 \\
        &
        &&
        \\
\scalebox{1}{  \begin{tikzpicture}[
roundnode/.style={circle, draw=black!60, fill=green!5},
squarednode/.style={rectangle, draw=black!60, fill=green!5},
]
  \begin{scope}[shift={(0, 0)}]


\node[roundnode] at (1,1) (loc) {\(l_{14}\)};
\node[squarednode,rounded corners]     at  (1,0)      (ret)                            {\(\var{ret}\)};

\draw[thick, -latex] (loc.south) -- (ret.north);
  \end{scope}
\end{tikzpicture}}
      &
         \raisebox{0\height}{ \ \scalebox{1}{  \begin{tikzpicture}[
roundnode/.style={circle, draw=black, fill=red!5},
squarednode/.style={rectangle, draw=black, fill=red!5},
]
  \begin{scope}[shift={(0, 0)}]

\node[squarednode,rounded corners]   at  (1,3)    (par)                              {\(\var{par}_{2}\)};
\node[squarednode,rounded corners]   at  (1,2)    (ret)                              {\(\var{ret}\)};
\node[roundnode] at (0,3) (loc) {\(l_{14}\)};
\draw[-latex,thick] (par.south) -- (ret.north);
\draw[-latex,thick] (loc.south) -- (ret.north);
  \end{scope}
\end{tikzpicture}}}
         &
         \raisebox{0\height}{          \scalebox{1}{  \begin{tikzpicture}[
roundnode/.style={circle, draw=black, fill=cyan!5},
squarednode/.style={rectangle, draw=black, fill=cyan!5},
]
  \begin{scope}[shift={(0, 0)}]

\node[squarednode,rounded corners]   at  (1,3)    (par)                              {\(\var{par}_{1}\)};
\node[squarednode,rounded corners]   at  (1,2)    (ret)                              {\(\var{ret}\)};
\node[roundnode] at (0,3) (loc) {\(l_{14}\)};
\draw[-latex,thick] (par.south) -- (ret.north);
\draw[-latex,thick] (loc.south) -- (ret.north);
  \end{scope}
\end{tikzpicture}}
         }
         &
         \\
        &
        &&
        \\
    (d) \prog{polyZ}()
    &
    (e) \prog{foo}(), \prog{mid}()
    &
    (f) \prog{bar}1(), \prog{bar}2()
    &
    (g) \prog{foo}()
    \\
  \end{tabular}
  \caption{Procedure summaries from a compositional and context insensitive analysis.}
  \label{fig:overCFCI}
\end{figure}
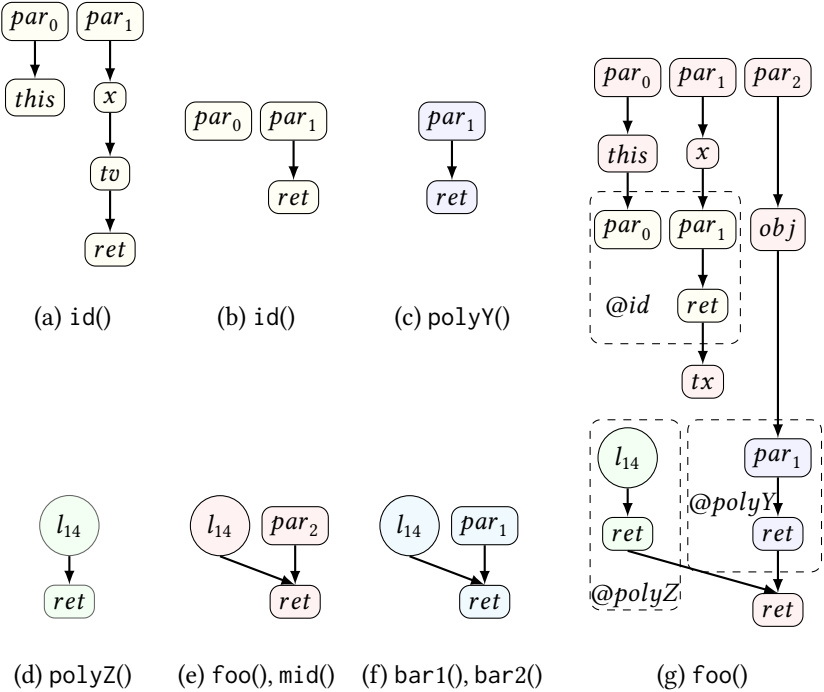

\emph{Summarization}. 
We introduce additional symbolic variables to denote the return value and the formal parameters of each procedure. For instance, the body of the procedure \prog{id}() at lines 18--22 is rewritten to ``\prog{this}=\(\mathtt{par}_0\)@\prog{id};\prog{X}\;\prog{x} = \(\mathtt{par}_1\)@\prog{id};\prog{X}\; \prog{tv} = \prog{x}; \prog{ret}@\prog{id} = \prog{tv}'', where \(\mathtt{par}_0\)@\prog{id} denotes its \(0\)th parameter and  \prog{ret}@\prog{id} denotes its return value. The summarization on these assignment statements is classic. On the assignment \prog{X}\; \prog{tv} = \prog{x} at line 19, the analysis generates the summary \(\forall \absEleTop \in \absEleTopSet, \absEleTop \mapsto \absEleTop \cup \{tv \supseteq x\} \), which indicates that for any precondition, the statement at line 19 just adds a new constraint \(tv \supseteq x\). The summary of \prog{id}() is obtained after applying summarization to the four statements as \(\forall \absEleTop \in \absEleTopSet, \absEleTop\mapsto \absEleTop\cup \{ret@\var{id} \supseteq tv, tv\supseteq x,x \supseteq par_1@\var{id},this \supseteq par_0@\var{id}\}\). Actually, all abstract summaries are of the form \(\forall \absEleTop \in \absEleTopSet, \absEleTop \mapsto \absEleTop \cup \absEleTop'\), which means that each summary just adds new constraints \(\absEleTop'\) to any precondition. In the following, we will only show the new constraints in each summary for simplicity. 
We also denote such constraints with  a graphical representation as shown in Figure~\ref{fig:overCFCI}(a) where``\(\leftarrow\)'' represents ``\(\supseteq\)''. Note that, we omit the label @\prog{id} in the figure if the procedure name is obvious. In these constraints, some variables like \prog{this}, \prog{tv} and \prog{x} cannot be accessed outside the procedure \prog{id}(). It is not necessary to include them in the summary. Thus  a transitive closure is computed and the constraints on the three variables are safely removed. The resulting summary is \( \{ret@\var{id} \supseteq par_1@\var{id}\}\) as shown in Figure~\ref{fig:overCFCI}(b), which indicates an identity procedure. Following the same principle, the summary of \prog{polyY}() (resp. \prog{polyZ}()) from class \prog{Y} (resp. \prog{Z}) is found as shown in Figure~\ref{fig:overCFCI}(c) (resp. Figure~\ref{fig:overCFCI}(d)). In these figures, \(l_{14}\) denotes the allocation site at line 14. We also omit some irrelevant nodes like \(\mathtt{par}_0@\mathtt{polyY}\).


In this compositional and context insensitive analysis, the summaries of the called procedures are inlined at each callsite. For instance, on the procedure invocation  at line 24 (\ie, \prog{X} \;\prog{tx} = \prog{id(x)}), the analysis inlines  the summary of \prog{id}(). The result is shown in Figure~\ref{fig:overCFCI}(g).
 The analysis loses no precision for this particular inlining. However, it will lose precision on the following statement \(\mathtt{ret}\)@\prog{foo}=\prog{tx.poly(obj)} at line 25 due to context insensitivity.

In the statement \(\mathtt{ret}\)@\prog{foo}=\prog{tx.poly(obj)}, the implementation of the abstract procedure \prog{poly}() is only known when the points-to set of its receiver (the variable \prog{tx}) is decided. 
The analysis can infer that the points-to set of the receiver depends on the first parameter of the procedure (\ie, \(\var{tx} \supseteq \var{par}_{1}@\var{foo}\) as shown in  Figure~\ref{fig:overCFCI}(g)). 
This means that the choice on the implementation of \prog{poly}() depends  on the context, which is absent at this stage. The compositional analysis  has to assume all possible input states where \(\mathtt{par}_{1}@\mathtt{foo}\) can point to any allocation site. Thus, the analysis has to enumerate all implementations of \prog{poly}() on this procedure invocation~\footnote{we do not take the error state into account for simplicity}. Thus, the summaries of \prog{polyY}() in Figure~\ref{fig:overCFCI}(c) and  \prog{polyZ}() in Figure~\ref{fig:overCFCI}(d) are both inlined, as shown in Figure~\ref{fig:overCFCI}(g). This brings precision loss. Note that we omit some irrelevant constraints like \(par_{0}@\var{polyY} \supseteq tx\). 

 The summary of \prog{foo}() is obtained as in Figure~\ref{fig:overCFCI}(e) after removing the inaccessible variables (line 23 is assumed to  contain irrelevant code). 
 The connection between the context and the choice on the implementation of \prog{poly}() is lost. 
 Similarly, the summary of \prog{mid}() is also obtained as in Figure~\ref{fig:overCFCI}(e). 
 Inlining this summary at the callsite at line 31 (\ie \prog{mid(new Y(), obj}) will produce the summary of \prog{bar}1() in Figure~\ref{fig:overCFCI}(f), which imprecisely includes \(l_{14}\) in the return value. Following the same principle, the summary of \prog{bar}2() also is obtained  as Figure~\ref{fig:overCFCI}(f).  They are less precise than inlining the statements of \prog{foo}() and \prog{mid}() at each callsite. Due to this imprecision, the analysis fails to verify either of the two assertions in   \prog{service}(). Worse still, this pointer analysis also generates a call graph that imprecisely includes the unreachable call string \prog{bar}1.\prog{foo}.\prog{polyZ}.

One way to achieve context sensitivity on conventional compositional analysis is via disjunctive summaries. However,  disjunction is known to be hard to manage and the size of which can grow rapidly~\cite{li2017semantic,zhang2014hybrid}. We will illustrate this later in this section.


\subsection{\toolname}
\label{sec:overviewinling}
Now we show how to generate a compositional and  context sensitive analysis with \toolname.

\emph{Hybrid summary}. We use \(\syntStmt \in \syntStmtset\) to denote program statements. We define a \emph{hybrid summary} as \( (\absEleCom, \critset)\), where \(\critset \in 2^\syntStmtset\) denotes a set of critical statements to be propagated and  \(\absEleCom \in \absEleTopSet \times \absEleTopSet\) is an abstract summary from the aforementioned compositional pointer analysis. A formal definition of critical statements will be given in Section~\ref{sec:formalization}. Here, we just give an intuitive introduction with the code in Figure~\ref{fig:motivating}. The statement \prog{X} \prog{tv=x} at line 19 is non-critical because no matter what the precondition is, the postcondition is always the constraint \(tv \supseteq x\). However, the virtual invocation statement \prog{tx.poly(obj)} at line 25 is critical because different preconditions can generate different postconditions (causing different procedures to be dispatched).

\emph{Summarization and inlining}. Our approach produces one  summary for  \prog{id}(), the same as in Figure~\ref{fig:overCFCI}(b) and inlines it at line 24 (\ie, \prog{X} \;\prog{tx} = \prog{id(x)}). However, on summarizing   \prog{foo}(), the situation is different. Our approach finds that summarizing the critical virtual invocation statement \prog{tx}.\prog{poly}(\prog{obj}) without any context will cause precision loss, and will thus add it to 
\( \critset\) instead of summarizing it immediately. The resulting hybrid summary of \prog{foo}() is 
\((\{tx \supseteq par_1@foo,obj \supseteq par_2@foo, ret \supseteq ret@poly\},\{\prog{tx}.\prog{poly}(\prog{obj})\}) \), as shown in Figure~\ref{fig:overCFCS}(a). The thick box denotes the critical statement and is connected with all variables it accesses. In our approach, critical statements are just like placeholders in the hybrid summary and will be summarized when the context is ready.

\begin{figure}[ht]
  \centering
  \begin{tabular}{  c  c  c}
             \scalebox{1}{  \begin{tikzpicture}[
roundnode/.style={circle, draw=black, fill=green!5},
squarednode/.style={rectangle, draw=black, fill=red!5},
squaredid/.style={rectangle, draw=black, fill=yellow!5},
squaredpolyy/.style={rectangle, draw=black, fill=blue!5},
squaredpolyz/.style={rectangle, draw=black, fill=green!5},
]
  \begin{scope}[shift={(0, 0)}]

\node[squarednode,rounded corners]   at  (1,0)    (parfoo2)                              {\(\var{par}_{2}\)};
\node[squarednode,rounded corners]   at  (0,0)    (parfoo1)                              {\(\var{par}_{1}\)};
\node[squarednode,rounded corners]   at  (1,-1)    (obj)                              {\(\var{obj}\)};
\node[squarednode,rounded corners]   at  (0,-1)    (tx)                              {\(\var{tx}\)};

\node[squarednode,ultra thick, rounded corners]   at  (0.5,-2)    (stmt)                              {\(\var{tx.poly(obj)}\)};
\node[squarednode,rounded corners]   at  (-1,-3)    (ret)                              {\(\var{ret}\)};

\node[squarednode,rounded corners]   at  (0.5,-3)    (retpoly)                              {\(\var{ret}@\var{poly}\)};
\draw[thick,-latex] (parfoo2.south) -- (obj.north);
\draw[thick,-latex] (parfoo1.south) -- (tx.north);
\draw[thick] (tx.south) -- (0,-1.7);
\draw[thick] (obj.south) -- (1,-1.7);
\draw[thick] (0.5,-2.3) -- (retpoly.north);
\draw[thick,-latex] (retpoly.west) -- (ret.east);
  \end{scope}
  \end{tikzpicture}} &
         \scalebox{1}{  \begin{tikzpicture}[
roundnode/.style={circle, draw=black, fill=green!5},
squarednode/.style={rectangle, draw=black, fill=red!5},
squaredid/.style={rectangle, draw=black, fill=yellow!5},
squaredpolyy/.style={rectangle, draw=black, fill=blue!5},
squaremid/.style={rectangle, draw=black, fill=green!5},
]
  \begin{scope}[shift={(0, 0)}]

\node[squaremid,rounded corners]   at  (1.3,3)    (parmid2)                              {\(\var{par}_{2}\)};
\node[squaremid,rounded corners]   at  (-0.3,3)    (parmid1)                              {\(\var{par}_{1}\)};

\node[squarednode,rounded corners]   at  (1.3,2)    (obj)                              {\(\var{obj}@\var{foo}\)};
\node[squarednode,rounded corners]   at  (-0.3,2)    (tx)                              {\(\var{tx}@\var{foo}\)};
 
\node[squarednode,ultra thick, rounded corners]   at  (0.5,1)    (stmt)                              {\(\var{tx.poly(obj)}\)};
\node[squaremid,rounded corners]   at  (-1,0)    (retfoo)                              {\(\var{ret}\)};

\node[squarednode,rounded corners]   at  (0.5,0)    (retpoly)                              {\(\var{ret}@\var{poly}\)};
\draw[thick,-latex] (parmid2.south) -- (obj.north);
\draw[thick,-latex] (parmid1.south) -- (tx.north);

\draw[thick] (tx.south) -- (0,1.3);
\draw[thick] (obj.south) -- (1,1.3);
\draw[thick] (0.5,0.7) -- (retpoly.north);
\draw[thick,-latex] (retpoly.west) -- (retfoo.east);
  \end{scope}
  \end{tikzpicture}} &
              \raisebox{0.7\height}{ \scalebox{1}{  \begin{tikzpicture}[
roundnode/.style={circle, draw=green!60, fill=green!5},
squarednode/.style={rectangle, draw=black, fill=cyan!5},
]
  \begin{scope}[shift={(0, 0)}]

\node[squarednode, rounded corners ]   at  (1,1)    (par)                              {\(\var{par}_{1}\)};
\node[squarednode, rounded corners]     at  (1,0)      (ret)                            {\(\var{ret}\)};
\draw[-latex,thick] (par.south) -- (ret.north);
  \end{scope}
\end{tikzpicture}}}
              \\
      (a) \prog{foo}()
      &
      (b) \prog{mid}()
      &
      (c) \prog{bar}1() \#3
      \\
            &\\
\scalebox{1}{  \begin{tikzpicture}[
roundnode/.style={circle, draw=black, fill=cyan!5},
squarednode/.style={rectangle, draw=black, fill=green!5},
squaredfoo/.style={rectangle, draw=black, fill=red!5},
squaredpolyy/.style={rectangle, draw=black, fill=blue!5},
squarebar/.style={rectangle, draw=black, fill=cyan!5},
]
  \begin{scope}[shift={(0, 0)}]

\node[squarebar,rounded corners]   at  (1.3,5)    (parbar1)                              {\(\var{par}_{1}\)};
\node[roundnode,rounded corners]   at  (-0.3,5)    (loc)                              {\(\var{l}_{31}\)};

\node[squarebar,rounded corners]   at  (1.3,4)    (midobj)                              {\(\var{obj}\)};

\node[squarednode,rounded corners]   at  (1.3,3)    (parmid2)                              {\(\var{par}_{2}\)};
\node[squarednode,rounded corners]   at  (-0.3,3)    (parmid1)                              {\(\var{par}_{1}\)};
\node[squaredfoo,rounded corners]   at  (1.3, 2)    (obj)                              {\(\var{obj@foo}\)};
\node[squaredfoo,rounded corners]   at  (-0.3,2)    (tx)                              {\(\var{tx@foo}\)};
 
\node[squaredfoo,ultra thick, rounded corners]   at  (0.5,1)    (stmt)                              {\(\var{tx.poly(obj)}\)};
\node[squarednode,rounded corners]   at  (-1,0)    (retbar)                              {\(\var{ret}\)};

\draw[rounded corners,dashed] (-1.5,-0.5) rectangle (2.2,3.5);
\tkzputform{-0.9,2.5}{@\var{mid}}

\node[squarebar,rounded corners]   at  (-1,-1)    (ret)                              {\(\var{ret}\)};
\node[squaredfoo,rounded corners]   at  (0.5,0)    (retpoly)                              {\(\var{ret}@\var{poly}\)};
\draw[thick,-latex] (parbar1.south) -- (midobj.north);

\draw[thick,-latex] (midobj.south) -- (parmid2.north);
\draw[thick,-latex] (loc.south) -- (parmid1.north);

\draw[thick,-latex] (parmid2.south) -- (obj.north);
\draw[thick,-latex] (parmid1.south) -- (tx.north);
\draw[thick] (tx.south) -- (0,1.3);
\draw[thick] (obj.south) -- (1,1.3);
\draw[thick] (0.5,0.7) -- (retpoly.north);
\draw[thick,-latex] (retpoly.west) -- (retbar.east);
\draw[thick,-latex] (retbar.south) -- (ret.north);
  \end{scope}
  \end{tikzpicture}} &
\scalebox{1}{  \begin{tikzpicture}[
roundnode/.style={circle, draw=black, fill=cyan!5},
squarednode/.style={rectangle, draw=black, fill=green!5},
squaredfoo/.style={rectangle, draw=black, fill=red!5},
squaredpolyy/.style={rectangle, draw=black, fill=blue!5},
squarebar/.style={rectangle, draw=black, fill=cyan!5},
]
  \begin{scope}[shift={(0, 0)}]

\node[squarebar,rounded corners]   at  (1.3,5)    (parbar1)                              {\(\var{par}_{1}\)};
\node[roundnode,rounded corners]   at  (-0.3,5)    (loc)                              {\(\var{l}_{31}\)};

\node[squarebar,rounded corners]   at  (1.3,4)    (midobj)                              {\(\var{obj}\)};

\node[squarednode,rounded corners]   at  (1.3,3)    (parmid2)                              {\(\var{par}_{2}\)};
\node[squarednode,rounded corners]   at  (-0.3,3)    (parmid1)                              {\(\var{par}_{1}\)};
\node[squaredfoo,rounded corners]   at  (1.3, 2)    (obj)                              {\(\var{obj@foo}\)};
\node[squaredfoo,rounded corners]   at  (-0.3,2)    (tx)                              {\(\var{tx@foo}\)};
 
\node[squaredpolyy,rounded corners]   at  (0.5,1)    (parpoly1)                              {\(\var{par}_1@\var{polyY}\)};

\node[squarednode,rounded corners]   at  (-1,0)    (retbar)                              {\(\var{ret}\)};

\draw[rounded corners,dashed] (-1.5,-0.5) rectangle (2.2,3.5);
\tkzputform{-0.9,2.5}{@\var{mid}}

\node[squarebar,rounded corners]   at  (-1,-1)    (ret)                              {\(\var{ret}\)};
\node[squaredpolyy,rounded corners]   at  (0.5,0)    (retpoly)                              {\(\var{ret}@\var{polyY}\)};
\draw[thick,-latex] (parbar1.south) -- (midobj.north);

\draw[thick,-latex] (midobj.south) -- (parmid2.north);
\draw[thick,-latex] (loc.south) -- (parmid1.north);

\draw[thick,-latex] (parmid2.south) -- (obj.north);
\draw[thick,-latex] (parmid1.south) -- (tx.north);
\draw[thick,-latex] (obj.south) -- (parpoly1.north);
\draw[thick,-latex] (parpoly1.south) -- (retpoly.north);
\draw[thick,-latex] (retpoly.west) -- (retbar.east);
\draw[thick,-latex] (retbar.south) -- (ret.north);
  \end{scope}
  \end{tikzpicture}} &
                \raisebox{0.7\height}{ \scalebox{1}{  \begin{tikzpicture}[
roundnode/.style={circle, draw=black!60, fill=green!5},
squarednode/.style={rectangle, draw=black!60, fill=green!5},
]
  \begin{scope}[shift={(0, 0)}]


\node[roundnode] at (1,1) (loc) {\(l_{14}\)};
\node[squarednode,rounded corners]     at  (1,0)      (ret)                            {\(\var{ret}\)};

\draw[thick, -latex] (loc.south) -- (ret.north);
  \end{scope}
\end{tikzpicture}}} \\\\
        (d) \prog{bar}1() \#1  
        & 
        (e) \prog{bar}1() \#2
        &
        (f) \prog{bar}2() \\
  \end{tabular}
  \caption{Procedure summaries via \toolname.}
  \label{fig:overCFCS}
\end{figure}

After inlining the hybrid summary of \prog{foo}() and removing the inaccessible variables, our analysis can obtain the hybrid summary of \prog{mid}() in Figure~\ref{fig:overCFCS}(b). Inlining this  hybrid  summary at line 31 in the procedure \prog{bar}1(), our analysis produces its hybrid summary in Figure~\ref{fig:overCFCS}(d). During this process, the critical statement \prog{tx}.\prog{poly}(\prog{obj}) is propagated from \prog{foo}() to \prog{bar}1().

\emph{Context ready predicate}. We introduce a \emph{context ready} predicate \(\predcrit\), which decides whether a critical statement should be summarized with the current context. A simple way to define it is to summarize critical statements that propagate \(k\) steps, i.e., callsites,   giving rise to \(k\)-callsite sensitivity. But we prefer to have a better definition for \(\predcrit\). In Figure~\ref{fig:overCFCS}(d), the receiver of the virtual invocation statement  \prog{tx}.\prog{poly}(\prog{obj}) is known to point to \(l_{31}\) (\ie the allocation of an instance of \prog{Y} at line 31). Our analysis also infers that the points-to set of the receiver \prog{tx} will not change even the analysis propagates this virtual invocation statement further, because  \prog{tx} and all its aliased variables are not accessible out of the scope of \prog{bar}1(). This context is considered as an \emph{adequate} context. We  define \(\predcrit\) as checking whether the context is adequate. Thus, the propagation of  \prog{tx}.\prog{poly}(\prog{obj}) stops and our analysis summarizes it with the current context. In this case, our analysis dispatches the virtual invocation to the implementation of \prog{poly}() in \prog{Y} (\ie, \prog{polyY}()). Then our analysis inlines the summary of \prog{polyY}() as shown in Figure~\ref{fig:overCFCS}(e). After removing the inaccessible variables, the summary of the procedure \prog{bar}1() is generated as shown in Figure~\ref{fig:overCFCS}(c). Similarly, the summary of   \prog{bar}2() is computed as shown in  Figure~\ref{fig:overCFCS}(f).

 With the summaries of  \prog{bar}1() and  \prog{bar}2() computed by our analysis, the two assertions at lines 40 and 41 can be verified successfully. 
In addition, the unreachable call string \prog{bar}1.\prog{foo}.\prog{polyZ} that is introduced
by the conventional compositional analysis
is ruled out in our precisely built call graph.
 

\emph{Comparison with conventional compositional approaches}. 
Compared with the conventional compositional context insensitive  analysis, \toolname achieves context sensitivity at some additional performance overhead on analyzing the critical statement \prog{tx}.\prog{foo}(\prog{poly}) twice in this example. This is a very good trade-off, especially, if most statements in a program are non-critical. Fortunately, this is the case as validated in our experiments, as described in Section~\ref{sec:evaluaiton}.

Another way to make compositional analysis context sensitive is via disjunction. Many approaches e.g., multiple summaries and disjunctive abstract states~\cite{dillig2011precise,feng2015bottom,zhang2014hybrid} fall into this category.
With this approach,  \(\prog{tx}.\prog{poly}(\prog{obj})\) is summarized as:
\begin{center}
    \(
    \begin{array}{c}
    \text{ if } \{tx \supseteq \{l_{Y}\}\} \in \absEleTop , \absEleTop \mapsto \absEleTop \cup \{ret@foo \supseteq \var{par}_{1}@\var{foo}\}  \\
        \text{ if } \{tx \supseteq \{l_{Z}\}\} \in \absEleTop, \absEleTop \mapsto \absEleTop \cup \{ret@foo \supseteq \{l_{14}\}\}  \\
    \end{array}
    \)
\end{center}
where
 \(l_{Y}\) (resp. \(l_{Z}\)) denotes an allocation site of object \prog{Y} (resp. \prog{Z}). This summary can achieve the same precision as \toolname. However, the main limitation~\cite{li2017semantic,zhang2014hybrid} of these approaches is well-known. 
The size of disjuncts increases exponentially  in terms of the number of virtual invocation statements. Suppose that class \prog{X} has a total of 10 different implementations and the procedure \prog{foo}() is added one parameter that is used as the receiver of another virtual invocation of \prog{poly}(). Then these approaches will end up building as many as 100 disjuncts in order to be equivalently precise  as \toolname. It takes much time and memory to generate and maintain the disjuncts. However, Hybrid Inlining just needs to propagate only two critical statements in this case.
Sometimes, it is just impossible to generate disjunction, because the number of possibilities are unbounded. Section~\ref{sec:pointer} shows such an example.


\emph{Comparison with conventional top-down approaches}.
The two assertions in Figure~\ref{fig:motivating} can be non-trivial for top-down Andersen-style pointer analysis to verify. It needs at least a 3-callsite-sensitive analysis. Such an analysis has to analyze the procedures \prog{foo}(), \prog{id}() and \prog{mid}() twice. This brings significant performance loss. Our approach just needs to analyze the statement at line 25 twice. All other statements are summarized once and for all.  Selective context sensitivity for top-down analysis does not help on this example, because all procedures need context sensitivity. What is worse, a \(k\)-object-sensitive analysis would simply fail here no matter what value \(k\) is given~\cite{jeon2022return}. 

Some CFL-based inter-procedural analyses~\cite{reps1995precise,reps2005weighted,spath2019context} are top-down but also utilize summaries to boost performance. Actually, the main idea of the compositional and context insensitive analysis in Section~\ref{sec:overviewci} is very similar to the summarization process in~\cite{reps1995precise}. However,  it does not delay the summarization of critical statements, thus cannot avoid generating multiple summaries for the same procedure (like \(\prog{mid}\)() and \(\prog{foo}\)()). This brings serious performance loss.

\presec
\section{Methodology} \label{sec:formalization}
\postsec

In this section, we formalize the main idea of our framework. We first consider an application of \toolname on flow sensitive analysis and then move to  flow insensitive analysis.

We define our formalization on a simple yet general language:

\begin{center}
\(
\syntProg  :=  \syntStmt  \mid \syntProg \circ \syntProg  \mid \syntProg + \syntProg \mid \syntProg^{*}
\)
\end{center}

In this language, a program \(\syntProg\) can be an atomic statement  \(\syntStmt\) (including assignments, condition tests, procedure invocations, etc.), a non-deterministic choice  \(\syntProg+\syntProg\), a sequential composition  \(\syntProg \circ \syntProg \) or an iteration \(\syntProg^{*}\). We  define the set of concrete states as \(\concSet\) and the concrete semantics of a program  \(\syntProg\) as \(\cSem{\syntProg}:2^\concSet \Rightarrow 2^\concSet \). We denote the set of all statements as \(\syntStmtset\). We do not include \emph{store}s and \emph{load}s in this language since our framework can be applied on any analysis regardless of field sensitivity. In Section~\ref{sec:instantiation}, we will instantiate our framework to a field sensitive pointer analysis.

\subsection{\toolname on Flow Sensitive Analysis}

We first formalize a general top-down analysis and a general compositional analysis, and then show to combine the ideas behind them with \toolname.

\subsubsection{A Top-down Flow Sensitive Analysis}
\label{tdfs}
This can be characterized by a set of  abstract states \(\absEleTop \in \absEleTopSet\), an abstract transfer function \(\transTop: (\syntStmtset \times  \absEleTopSet) \Rightarrow \absEleTopSet) \) on atomic statements,  and a join operator \(\topJoin: \absEleTopSet \times \absEleTopSet \Rightarrow \absEleTopSet\).
An abstract state \(\absEleTop \in \absEleTopSet\) is connected to a set of concrete states via a concretization function \(\gamma: \absEleTopSet \Rightarrow 2^\concSet\).  Two special abstract states are \emph{top} \(\top \in \absEleTopSet\) and \emph{bottom} \(\bot \in \absEleTopSet\) 
 such that \(\gamma(\top) = \concSet\) and \(\gamma(\bot) = \emptyset\).  The abstract transfer function \(\transTop(\syntStmt)\) over-approximates the concrete semantics on statements:
\(
\forall \absEleTop \in \absEleTopSet, \gamma(\transTop(\syntStmt)(\absEleTop)) \supseteq 
\cSem{\syntStmt}(\gamma(\absEleTop))\). The join operator over-approximates the union of the two sets of concrete states: \(\gamma(\absEleTop_0\topJoin\absEleTop_1) \supseteq \gamma(\absEleTop_0)\cup \gamma(\absEleTop_1)\).

The abstract semantics \(\absSemTop{\syntProg}: \cieleset \Rightarrow \cieleset\) of a top-down flow sensitive analysis can be defined as:
\begin{center}
\(
\begin{array}{rcl}
    \absSemTop{\syntStmt} (\absEleTop) & := & \transTop(\syntStmt) (\absEleTop) \\
    \absSemTop{\syntProg_0 \circ \syntProg_1}(\absEleTop) & := & \absSemTop{\syntProg_1} (\absSemTop{\syntProg_0}( \absEleTop )) \\
    \absSemTop{\syntProg_0 +  \syntProg_1}( \absEleTop ) &:= & \absSemTop{\syntProg_0}(\absEleTop)\topJoin \absSemTop{\syntProg_1} (\absEleTop)\\
      \absSemTop{\syntProg^{*}}( \absEleTop ) &:= & \fixpoint_{\absEleTop}(\lambda \absEleTop'. (\absEleTop'\topJoin  \absSemTop{\syntProg}(\absEleTop')))\\
\end{array}
\)
\end{center}

This definition of top-down flow sensitive is standard. 
Here, \(\fixpoint_{\absEleTop}\) denotes a process for computing the least fix-point with \(\absEleTop\) as the initial state. Its soundness is obvious such that

\begin{center}
    \(
    \gamma(\absSemTop{\syntProg} (\absEleTop)) \supseteq \cSem{\syntProg}(\gamma(\absEleTop))
    \)
\end{center}

\begin{example}
A popular flow sensitive analysis is the Polyhedra domain~\cite{cousot1978automatic}. Its abstract states are defined as linear inequalities on variables, such as:
\[a_0 + \sum_{i} a_{i}x_{i} \leq 0\]
where \(a_{i}\) represents a constant and \(x_i\) represents a program variable. For the transfer functions and the join operator in the Polyhedra domain, we refer  to~\cite{cousot1978automatic}
\end{example}

\subsubsection{Strict Context Sensitivity}

We define strict context sensitivity on procedure invocation statements \(\procCall\). We write \(\mapproctostmt: \syntax{proc} \mapsto \syntProg\) to map a procedure \(\syntax{proc}\) to its statements. The abstract semantics is strictly context sensitive if for all procedure invocation statements, we have: 

\begin{center}

\(
\forall \absEleTop \in \absEleTopSet, \transTop(\syntax{call\;proc}) (\absEleTop) = \absSemTop{\mapproctostmt(\syntax{proc})} (\absEleTop)
\)

\end{center}

This requires the transfer function of an invocation statement to be as precise as re-analyzing the body of the procedure for each distinct precondition. In top-down analysis, this is almost equivalent  to analyzing a fully inlined program. Thus, even though strict context sensitivity is essential for precision, it is hardly practical due to serious performance concerns. 

\subsubsection{Compositional Flow Sensitive Analysis}
\label{bufs}
Unlike top-down analysis, compositional analysis is built on abstract summaries rather than abstract states. Abstract summaries are usually relations on abstract states mapping  preconditions to postconditions. A conventional summary can be defined as \(\absEleCom \in \absEleComSet: \absEleTopSet \Rightarrow \absEleTopSet\). A conventional compositional analysis is characterized by the following operators:

\begin{itemize}
    \item The abstract transfer function \(\transCom: (\absEleTopSet \times \syntStmtset) \Rightarrow \absEleComSet\) summarizes a statement in \(\syntStmtset\) to a summary in \(\absEleComSet\) with a context in \(\absEleTopSet\). The context represents a condition that must be satisfied by the precondition;
    \item A compositional operator \(\compose: (\absEleTopSet \times \absEleComSet \times \absEleComSet) \Rightarrow \absEleComSet\)  summarizes the effect of composition operation "\(\circ\)"; and
    \item A join Operator \(\comJoin: (\absEleTopSet \times \absEleComSet \times \absEleComSet) \Rightarrow \absEleComSet\) summarizes the effect of non-deterministic choice "\(+\)".

\end{itemize}

A compositional analysis first summarizes a procedure  as an abstract summary \(\absEleCom\) and then applies it on any precondition \(\absEleTop\) to compute the postcondition  \(\absEleCom(\absEleTop)\). In the summarization process, a more precise context can often lead to a more precise summary. For instance, for the abstract transfer function \(\transCom\), let \(\absEleTop,\absEleTop'\) be two contexts and \(\absEleTop''\)  a precondition, the following  always holds:
\begin{center}
    \(
    \forall \absEleTop,\absEleTop',\absEleTop'' \in \absEleTopSet, \gamma(\absEleTop)\supseteq \gamma(\absEleTop') \supseteq \gamma(\absEleTop'') \implies  \gamma(\transCom(\absEleTop,\syntStmt)(\absEleTop'')) \supseteq 
    \gamma(\transCom(\absEleTop',\syntStmt)(\absEleTop'')) 
    \)
\end{center}

As summarization with appropriate contexts can bring precision, a compositional analysis has to start with \(\top\) to fit into any precondition. We define the  summarization process \(\sumCom\) as follows:
\begin{center}
\(
\begin{array}{rcl}
    \sumCom(\syntStmt)  & := & \transCom(\top,\syntStmt)  \\
    \sumCom(\syntProg_0 \circ \syntProg_1) & := & \compose(\top,\sumCom(\syntProg_1), \sumCom(\syntProg_0)) \\
    \sumCom(\syntProg_0 +  \syntProg_1) &:= & \comJoin(\top,\sumCom(\syntProg_0), \sumCom(\syntProg_1)) \\
      \sumCom(\syntProg^{*}) &:= & \fixpoint_{\sumCom(\syntProg)}(\lambda \absEleCom.\comJoin(\top, \absEleCom,\compose(\top, \sumCom(\syntProg),\absEleCom)))\\
\end{array}
\)
\end{center}

With \(\sumCom\) (which is  straightforward), the abstract semantics of this compositional analysis is: 
\begin{center}
    \(
    \absSemCom{\syntProg} (\absEleTop) := \sumCom(\syntProg) (\absEleTop) 
    \)
\end{center}

This abstract semantics is  sound if the abstract operators are sound in the following sense:
\begin{center}
\(
\begin{array}{rl}
\forall \absEleTop \in \absEleTopSet, &\gamma (\transCom(\top,\syntStmt)(\absEleTop))\supseteq \gamma(\transTop(\syntStmt)(\absEleTop)) \supseteq \cSem{\syntStmt} \gamma(\absEleTop)
\\
\forall \absEleTop \in \absEleTopSet, \forall \absEleCom_0,\absEleCom_1\in \absEleComSet, &\gamma (\compose(\top,\absEleCom_1,\absEleCom_0)(\absEleTop)))\supseteq \gamma(\absEleCom_1(\absEleCom_0(\absEleTop)))
\\
\forall \absEleTop \in \absEleTopSet, \forall \absEleCom_0,\absEleCom_1\in \absEleComSet, &\gamma( \comJoin(\top,\absEleCom_0,\absEleCom_1)(\absEleTop))\supseteq \gamma(\absEleCom_0(\absEleTop)\topJoin\absEleCom_1(\absEleTop))
\end{array}
\)
\end{center}

We assume that \(\forall \absEleTop \in \absEleTopSet, \gamma (\transCom(\top, \syntStmt)(\absEleTop))\supseteq \gamma(\transTop(\syntStmt)(\absEleTop)) \),  indicating that the transfer function from top-down analysis is at least as precise as the corresponding compositional analysis. This is a valid assumption, because, otherwise, we can just define \(\transTop(\syntStmt)(\absEleTop)\) as \(\transCom(\top,\syntStmt)(\absEleTop)\). 

Recall  that \(\mapproctostmt: \syntax{proc} \mapsto \syntProg\)  maps a procedure \(\syntax{proc}\) to its statements. The transfer function  for procedure invocation is  defined to be:

\begin{center}

\( \sumCom(\procCall) := \sumCom(\mapproctostmt(\syntax{proc})) \).
    
\end{center}

From this definition, we can see that compared to top-down analysis, compositional analysis just needs to compute the summary \(\sumCom(\mapproctostmt(\syntax{proc}))\) of each procedure once and for all. This ensures that its analysis time  increases almost linearly as the size of the program
being analyzed. This represents a significant performance advantage of compositional analysis over top-down analysis.

Our definition  is a standard abstract relation based compositional analysis.


\begin{example}
A compositional version of the Polyhedra domain has been proposed in~\cite{cousot2002modular}. It uses linear inequalites (same as abstract states) to store summaries. To represent the relations of a preconditions and its corresponding postcondition, their  variables   are renamed differently.  For instance, to summarize a simple assignment \(\prog{x = x + 1}\), the summary is \(
x^{o} - x^{i} - 1 \leq 0 \wedge x^{i} - x^{o} - 1 \leq 0\), where \(x^{i}\) (resp. \(x^{o}\) represents a variable from the precondition (resp., postcondition)). The composition and join  operators in the compositional Polyhedra analysis just reuse the \emph{meet} operator and the join operator in the top-down Polyhedra domain, respectively. For more details, we refer to \cite{cousot1978automatic} and \cite{cousot2002modular}.

\end{example}

However, despite its performance advantage, compositional analysis  is much less explored than top-down analysis in practice. This is because it is difficult to design a compositional analysis that is as precise as its top-down counterpart, especially with respect to strict context sensitivity.
 
\subsubsection{Difficulties in Designing Strictly Context Sensitive Compositional Analysis}

Following the definition of strict context sensitivity, a compositional analysis is strict context sensitive iff:

\begin{center}
    \(
\forall  \absEleTop \in \absEleTopSet,  \sumCom(\mapproctostmt(\syntax{proc})) (\absEleTop) = \absSemTop{\mapproctostmt(\syntax{proc})}  (\absEleTop)
    \)
\end{center}

Therefore, for any precondition, applying the abstract summary of a procedure  will produce the same postcondition as re-analyzing its body. To achieve this, the transfer functions \(\transCom\), the composition operator \(\compose\) and the join operator \(\comJoin\) cannot lose precision. For any statement \(\syntStmt\) and abstract summaries \(\absEleCom_0,\absEleCom_1 \in \absEleComSet\),
the following three conditions must  be  thus satisfied:

\begin{center}
\(
\begin{array}{rcl}
\predicateSen_{\transCom}(\syntStmt)= true &\text{iff}& \forall \absEleTop \in \absEleTopSet, \transCom(\top,\syntStmt)(\absEleTop) =  \transTop(\syntStmt)(\absEleTop)
\\
\predicateSen_{\compose}(\absEleCom_0,\absEleCom_1)= true &\text{iff}& \forall \absEleTop \in \absEleTopSet, \compose(\top,\absEleCom_1,\absEleCom_0)(\absEleTop)) = \absEleCom_1(\absEleCom_0(\absEleTop))
\\
\predicateSen_{\comJoin}(\absEleCom_0,\absEleCom_1)= true &\text{iff}& \forall \absEleTop \in \absEleTopSet, \comJoin(\top,\absEleCom_0,\absEleCom_1)(\absEleTop) =\topJoin( \absEleCom_0(\absEleTop),\absEleCom_1(\absEleTop))
\end{array}
\)

\end{center}

It is very difficult (if not impossible) to design a compositional analysis satisfying all three conditions on non-trivial properties. There are two main difficulties: (1) a domain of computer representable abstract summaries (i.e., sets of symbolic relations) supporting composition and join is harder to design than abstract states in top-down analysis, and
(2) the behaviors of statements can be so complex that summarizing them without contexts can hardly keep all behaviors in the resulting summary. For instance, a compositional Polyhedra domain cannot precisely summarize any non-linear statements. In the case of a non-linear assignment \(\syntax{z = x * y}\), the compositional Polyhedra domain can only strongly update \(\syntax{z}\) by \((-\infty,+\infty)\). On the other hand, 
a top-down Polyhedra analysis can infer much more precise postcondition given a suitable precondition. If the precondition is \(\syntax{x} = 2 \wedge \syntax{y} = 2\), for example,
\(\syntax{z}\)  can be
  strongly updated to be  4.

Due to these two  difficulties (especially the second one),  compositional analysis is used relatively less often than   top-down analysis in practice. 
Conventional compositional analysis often gives up strict context sensitivity, or as in many prior works~\cite{dillig2011precise,feng2015bottom,zhang2014hybrid}, addressing the second difficulty with disjunctive summaries,  can easily give rise to  disjunctive explosion.

\subsubsection{\toolname for Flow Sensitive Analysis}

In this paper, we propose to  make conventional compositional analysis strictly context sensitive without introducing disjunction.  The approach is centered on one key observation: given a conventional compositional analysis, the predicate \(\predicateSen_{\transCom}(\syntStmt) \) evaluates to \(False\) only on some statements, which
are called \emph{critical} statements.

We say that a statement \(\syntStmt\) is \emph{critical}  if  \(\predicateSen_{\transCom}(\syntStmt) \)  does not hold:

\begin{center}
    \(
    \exists \absEleTop \in \absEleComSet, \transCom(\top,\syntStmt)(\absEleTop) \supset  \transTop(\syntStmt)(\absEleTop)
    \)
\end{center}

As mentioned earlier, the non-linear statements are \emph{critical} for the compositional Polyhedra domain. Similarly, we say that
a statement is 
\emph{non-critical}  if \(\predicateSen_{\transCom}(\syntStmt) \) holds:

\begin{center}
    \(
    \forall \absEleTop \in \absEleComSet, \transCom(\top,\syntStmt)(\absEleTop) =  \transTop(\syntStmt)(\absEleTop)
    \)
\end{center}

All linear assignments are   non-critical for the compositional Polyhedra domain. For instance, the statement \(\syntax{x = x + 3}\) can be summarized as \(0\leq x^{o}- x^{i} - 3\leq 0 \). Given any precondition, this summary can generate the same postcondition as the top-down Polyhedra domain.

With this observation, we propose to treat critical and non-critical statements differently during the summarization. Unlike  top-down analysis that inlines all statements of a callee and compositional analysis that inlines the summaries of all statements of the callee, our approach inlines all critical statements and the summaries of all non-critical statements in the callee. We achieve this with a 
\emph{hybrid summary} \(\absEleHybrid\), which is defined as follows:

\begin{center}
    
\(
\absEleHybrid  :=  \syntStmt \mid \absEleCom \mid \absEleHybrid \circ \absEleHybrid  \mid \absEleHybrid + \absEleHybrid \mid \absEleHybrid^{*}
\)
\end{center}

A hybrid summary is actually a complete procedure body with all non-critical statements summarized, computed with a function \(\sumHybrid\) that summarizes any \(\syntProg\) to a hybrid summary \(\absEleHybrid\).

On atomic statements, \(\sumHybrid\) just summarizes them  with \(\transCom\) if they are non-critical and keep them as they are if they are critical. This is defined as follows:
\begin{center}
    \(
      \sumHybrid(\syntStmt)   := 
    \left\{
\begin{array}{ll}
 \transCom(\top,\syntStmt), \qquad&if\; \predicateSen_{\transCom}(\syntStmt) \\
\syntStmt, \qquad & otherwise.
\end{array}
\right.
    \)
\end{center}

On program composition \(\syntProg_0 \circ \syntProg_1\), if both \(\syntProg_0\) and \(\syntProg_1\) can be summarized into abstract relations in \(\absEleComSet\) and \(\compose\) does not lose precision on them, they are summarized with \(\compose\). Otherwise, \(\sumHybrid\) just keeps this operation in the hybrid summary:


\begin{center}
    \(
     \sumHybrid(\syntProg_0 \circ \syntProg_1)  := 
    \left\{
\begin{array}{ll}
\compose(\top,\sumHybrid(\syntProg_1), \sumHybrid(\syntProg_0)), & if \begin{array}{l}
\sumHybrid(\syntProg_0)\in \absEleComSet \wedge \sumHybrid(\syntProg_1) \in \absEleComSet \\ \wedge
\predicateSen_{\compose}(\sumHybrid(\syntProg_0) ,\sumHybrid(\syntProg_1) )
\end{array}
\\
\sumHybrid(\syntProg_0)\circ \sumHybrid(\syntProg_1), & otherwise. \\
\end{array}
\right.
    \)
\end{center}

The summarization of non-deterministic choice and iteration follows the same principle:

\begin{center}
    \(
    \begin{array}{l}
      \sumHybrid(\syntProg_0 +  \syntProg_1)  := 
    \left\{
\begin{array}{ll}
\comJoin(\top,\sumHybrid(\syntProg_0), \sumHybrid(\syntProg_1)),& if \begin{array}{l}
\sumHybrid(\syntProg_0)\in \absEleComSet \wedge \sumHybrid(\syntProg_1) \in \absEleComSet \\\wedge
\predicateSen_{\comJoin}(\sumHybrid(\syntProg_0) ,\sumHybrid(\syntProg_1) )
\end{array}
\\
\sumHybrid(\syntProg_0) +  \sumHybrid(\syntProg_1),& otherwise. \\
\end{array}
\right.\\
      \sumHybrid(\syntProg^{*})  :=
    \left\{
\begin{array}{ll}
\fixpoint_{\sumHybrid(\syntProg)}(\lambda \absEleCom. \comJoin(\top,\absEleCom,\compose(\top,\absEleCom, \sumHybrid(\syntProg))), &if \begin{array}{l}
\sumHybrid(\syntProg)\in \absEleComSet
\wedge ( \forall \absEleCom \in \absEleComSet,\\
\predicateSen_{\comJoin}(\absEleCom, \compose(\top,\absEleCom, \sumHybrid(\syntProg))\\
\quad\wedge \predicateSen_{\compose}(\absEleCom, \sumHybrid(\syntProg))))
\end{array}
\\
\sumHybrid(\syntProg)^{*} ,  & otherwise. \\
\end{array}
\right.
\end{array}
    \)
\end{center}

Now we define the abstract semantics with hybrid inlining. Our hybrid summary contains both statements and  abstract summaries. As a result, we rely on an operator, \(\apply\) defined 
straightforwardly below, to compute the postcondition with respect to a given precondition:

\begin{center}
\(
\begin{array}{lcl}
    \apply(\syntStmt, \absEleTop)  & := & \absSemTop{\syntStmt}(\absEleTop)  \\
    \apply(\absEleCom, \absEleTop)  & := & \absEleCom(\absEleTop)  \\
    \apply(\absEleHybrid_0 \circ \absEleHybrid_1, \absEleTop) & := & 
    \apply(\absEleHybrid_0, \apply(\absEleHybrid_1, \absEleTop))
    \\
    \apply(\absEleHybrid_0 +  \absEleHybrid_1) (\absEleTop) &:= & \apply(\absEleHybrid_0, \absEleTop) \topJoin(\apply(\absEleHybrid_1, \absEleTop) \\
      \apply(\absEleHybrid^{*},\absEleTop) &:= & \fixpoint_{\absEleTop}(\lambda \absEleTop'.  (\absEleTop'\topJoin\apply(\absEleHybrid, \absEleTop')))\\
\end{array}
\)
\end{center}
where \(\apply\)   just applies abstract relations from the compositional domain and the abstract operators from  the top-down domain as needed. The abstract semantics with hybrid inlining is: 

\begin{center}
\(
\absSemHybrid{\syntProg}(\absEleTop):= \apply(\sumHybrid(\syntProg),\absEleTop)
\)
\end{center}

\begin{theorem}
The abstract semantics \(\absSemHybrid{\cdot} \) of the hybrid domain is sound if  the
abstract semantics for both the underlying compositional domain \(\absSemCom{\cdot} \)  and top-down domain \(\absSemTop{\cdot} \) are sound.
\end{theorem}

\begin{theorem}
The abstract semantics \(\absSemHybrid{\cdot} \) of the hybrid domain is strictly context sensitive.
\end{theorem}

\begin{example}
To apply \toolname to the Polyhedra domain, we just need to implement the three predicates:
\(\predicateSen_{\transCom}, \predicateSen_{\compose}\) and \(\predicateSen_{\comJoin}\). \(\predicateSen_{\transCom}\) evaluates all non-linear/linear statements to \(false\)/\(true\). This can be implemented 
with a simple syntactic check.   \(\predicateSen_{\transCom}\) is always \(true\), since the compositional operator (i.e., the \emph{meet} operator from top-down Polyhedra) loses no precision. \(\predicateSen_{\comJoin}\) evaluates to \(true\) when one operand represents a subset of concrete sets of the other. This can be implemented with the \emph{comparison} operator from ~\cite{cousot1978automatic}.

\end{example}

\subsubsection{Context Ready Predicate}
In the aforementioned algorithm, critical statements are propagated to the root procedure and abstracted with the top-down analysis.  However, this is not always necessary. We define the value set of a hybrid summary as \(\valueSet(\absEleHybrid) := \{ \absEleTop \mid \exists \absEleTop', \absEleTop = \apply(\absEleHybrid, \absEleTop')  \}\).  During the process of computing the summary \(\sumHybrid(\absEleHybrid \circ \syntStmt)\), the context of \(\syntStmt\) is actually \(\valueSet(\absEleHybrid)\). Thus,  \(\syntStmt\) can be re-summarized by \(\transCom(\valueSet(\absEleHybrid),\syntStmt) \).
As the hybrid summary grows in size,  the context \(\valueSet(\absEleHybrid)\) becomes more and more precise. For instance, the context \(\valueSet(\absEleHybrid_0 \circ \absEleHybrid_1)\) for \(\syntStmt\) in \(\absEleHybrid_0 \circ \absEleHybrid_1\circ\syntStmt\) is more precise than \(\valueSet(\absEleHybrid_1)\) in \(\absEleHybrid_1\circ\syntStmt\). 
 We define a \emph{context ready predicate}  \(\predcrit: \absEleTopSet \Rightarrow \{true, false\}\) which evaluates whether the current context is enough for the summarization of the critical statements. 
Our analysis can summarize the critical statements when the context is ready without propagating them to the root procedure. Thus, such  context ready predicates limit the propagation of critical statements. They can be used to trade-off between performance and precision. However, if we define a context ready predicate as being also \emph{adequate}, then no precision will be lost.  An adequate predicate should satisfy the following condition:

\begin{center}
\(
\textbf{if } \predcrit(\absEleTop) =true \textbf{ then }
\forall \absEleTop' \in \absEleTopSet, \gamma(\absEleTop')\subseteq \gamma(\absEleTop) \implies
\transCom(\absEleTop, \syntStmt)(\absEleTop') =  \transTop(\syntStmt)(\absEleTop')
\)
\end{center}

An adequate predicate evaluates to $true$ when the context  is precise enough so that \(\transCom\) loses no precision. As the critical statements propagate, their context  can only be more and more precise. Thus, our approach does not need to propagate them unnecessarily.
 For instance, the critical statement \(\syntax{z = x * y}\) can be safely summarized if the s statement is \(\syntax{x = 3}\) is encountered during the propagation. In this case \(\syntax{x = 3\circ z = x*y}\) is 
summarized precisely into \(x = 3 \wedge z = 3*y\). 

\begin{example}
In the compositional Polyhedra domain, there can be multiple adequate predicates  \(\predcrit\) for non-linear statements (e.g., \(x = y * z\)). There are two cases. 
One is that the variables in the right-hand side value expressions are confined by the same lower/upper bound (e.g., \(a\leq y \leq a\), where \(a\) is a constant). Another is that  the variables in the right-hand side value expressions are not accessible in the precondition and have no relational constraints with other accessible variables.
\end{example}

Our context ready predicates can be applied to the composition and join operators in the same way as  abstract transfer functions.

\begin{theorem}
Hybrid inlining with adequate predicates are sound and strictly context sensitive.
\end{theorem}

Adequate predicates are actually important for the performance of hybrid inlining. Many industrial applications can easily have  call stacks representing more than 100 callsites. Propagating all critical statements to the root procedures will cause unnecessary performance loss. In addition, hybrid inlining with adequate predicates induces a compositional analysis, rather than a hybrid top-down and compositional analysis, because it does not need the knowledge of the call string from the root procedures when summarizing a procedure. This makes it possible to reap the 
benefits of compositional analysis, by, e.g., parallelizing and incrementalizing our
analysis easily. However, sometimes there can be too many critical statements. We can  limit the maximal steps each critical  statement can propagate as \(k\) (similar to \(k\)-callsite-sensitivity in top-down analysis). We call it \(k\)-\emph{limit} Hybrid Inlining.

\subsubsection{Analysis Process and Recursion Handling}
\label{recursion}
Our analysis is simple, as it just summarizes all reachable procedures with \(\sumHybrid\) and then evaluates all the root procedures with \(\apply\). However, as revealed by the algorithm of \(\sumHybrid\), the summarization of a procedure needs to use the hybrid summaries in its called procedures. Since all procedures are analyzed in a random order (unlike in other works, our analysis does not build a call graph via a pre-analysis), it can happen that some callees are not summarized yet when a caller is to be summarized. In this case, our analysis just turns to summarize all the callees first,
by adopting a simple depth-first search algorithm. However, we need to specifically handle call string recursion to ensure the termination of our algorithm. During the depth-first search, our algorithm maintains the accumulated call string. Once recursion is detected on the call string,
our analysis will treat it by unrolling it with a factor of  $n$.

Our approach for handling recursion has two advantages: (1) it is simple to implement; and (2) its precision is comparable to an $(n-1)\times l$ callsite sensitive top-down analysis at each recursion site, where $l$ is the length of the recursive call string;  In practice, we set $n=2$ in order to strike a trade-off between precision and performance. 

\subsubsection{Application Scope and Limitations}

Our \toolname framework is general and can be applied to any top-down analysis within the scope of Section~\ref{tdfs} and any compositional analysis within the scope of Section~\ref{bufs} regardless of the underlying programming language used. All the technical definitions in Section~\ref{tdfs}~\cite{cousot1977abstract} and in Section~\ref{bufs}~\cite{sharir1978two} are classic. To  the best of our  knowledge, most popular top-down and compositional flow sensitive analyses fall into these categories. 

It is  easy to integrate a specific analysis with \toolname, which reuses the abstract operators from the underlying top-down analysis and compositional analysis. The additional work just involves defining the predicates \(\predicateSen_{\transCom}, \predicateSen_{\compose}\), \(\predicateSen_{\comJoin}\) and \(\predcrit\) for the underlying analysis, which determines whether a statement or an operator should be propagated or summarized.  Such predicates are easy to implement as  shown  for the non-trivial abstract domain Polyhedra. 

We do have one concern on \toolname for flow sensitive analysis: the memory occupation of hybrid summaries can be large. Consider a program with  three statements \(\syntStmt_0 \circ \syntStmt_1 \circ \syntStmt_2\). It can be summarized to \(\absEleCom_0 \circ \syntStmt_1 \circ \absEleCom_2\) when \(\syntStmt_1\) is critical. In many abstract domains, an abstract summary \(\absEleCom_0\) usually takes  much larger memory than \(\syntStmt_0\). If a program contains many critical statements, the memory occupation can become a serious concern. However, this concern can be adequately addressed in flow insensitive analysis, which will be described below.

\subsection{Flow Insensitive Analysis}
\label{sec:flow-insen}
Applying \toolname on flow insensitive analysis follows the same principle on flow sensitive analysis. For simplicity, we reuse most of the notations from those on flow sensitive analysis.

\subsubsection{Top-Down Flow-Insensitive Analysis}

This can be specified by the abstract transfer functions \(\transTop: (\syntStmtset \times \absEleTopSet) \Rightarrow \absEleTopSet \), which over-approximate the behaviors of program statements and a join operator \(\topJoin : \absEleTopSet \times \absEleTopSet \Rightarrow \absEleTopSet\) that over-approximates the union of sets of concrete states.
A flow insensitive analysis sees a procedure as a set of statements.
We define  a function \(\progToSet\) to map a program \(\syntProg\) to the set of statements \(\syntStmts\) contained in  \(\syntProg\).

The abstract semantics of the top-down flow insensitive analysis is defined as follows:
\[
\begin{array}{c}
\absSemTop{\syntProg} = \fixpoint_{\bot}(\lambda \absEleTop. \bigsqcup\nolimits^{t}\limits_{\syntStmt \in \progToSet(\syntProg)}\transTop(\syntStmt)(\absEleTop))
\end{array}
\]

This standard definition is sound if its transfer functions and join operator are sound. The formal definition of soundness is similar to that of flow sensitive analysis and thus omitted.

The strict context sensitivity on flow insensitive analysis is also similar. 
Recall that the function \(\mapproctostmt: \syntax{proc} \mapsto \syntProg\) maps a procedure \(\syntax{proc}\) to its statements. The abstract semantics is strictly context sensitive if the following condition holds for all procedure invocation statements:

\begin{center}

\(
\forall \absEleTop \in \absEleTopSet, \transTop(\syntax{call\;proc}) (\absEleTop) = \absSemTop{\mapproctostmt(\syntax{proc})} (\absEleTop)
\).
    
\end{center}

As in the case of flow sensitive analysis, a top-down flow insensitive analysis also needs a 
full inlining policy to achieve strict context sensitivity. This  brings serious performance concerns.

\subsubsection{Compositional and Flow Insensitive Analysis}
\label{compFI}
A conventional compositional and flow insensitive analysis can be characterized by a set of abstract summaries \(\absEleCom \in \absEleComSet: \absEleTopSet \Rightarrow \absEleTopSet\), an abstract transfer function \(\transCom: (\syntStmtset  \times \absEleTopSet) \Rightarrow \absEleComSet\), and a join Operator \(\comJoinI: ( \absEleComSet \times \absEleComSet) \Rightarrow \absEleComSet\). These operators are similar to those in flow sensitive analysis. The only  difference is that the join operator \(\comJoinI\) does not take any context as input any more. Instead, it always assumes the context to be \(\top\). We will see the reason for this later. Compositional analysis comes with a summarization function \(\sumCom\) that maps a program to an abstract relation, which is defined below:

\begin{center}
\(
    \sumCom(\syntProg)  :=  \bigsqcup^{c}_{\syntStmt \in \progToSet(\syntProg)}(\transCom(\top, \syntStmt) )
\)
\end{center}

The abstract semantics of this compositional analysis is defined as: 

\begin{center}
    \(
    \absSemCom{\syntProg} (\absEleTop) := \fixpoint_{\absEleTop}(\sumCom(\syntProg))
    \)
\end{center}

The summarization of procedure invocation statement is defined as:

\begin{center}

\( \sumCom(\procCall) := \sumCom(\mapproctostmt(\syntax{proc})) \).
    
\end{center}

 Compositional flow insensitive analysis  just needs to compute the summary \(\sumCom(\mapproctostmt(\syntax{proc}))\) for each procedure once and for all, by  sharing
 the same performance advantage as in the case of 
 compositional flow sensitive analysis.

\begin{example}
The Andersen-style pointer analysis discussed in Section~\ref{sec:overviewci} falls into the category of compositional, flow insensitive and context insensitive analysis defined in this section.
\end{example}

A compositional flow insensitive analysis is strictly context sensitive if the following holds:

\begin{center}
    \(
\forall  \absEleTop \in \absEleTopSet,  \sumCom(\mapproctostmt(\syntax{proc})) (\ciele) = \absSemTop{\mapproctostmt(\syntax{proc})}  (\ciele)
    \)
\end{center}

This condition holds if   for any statement \(\syntStmt\) and abstract summaries \(\absEleCom_0,\absEleCom_1 \in \absEleComSet\),
the following two conditions must be satisfied:

\begin{center}
\(
\predicateSen_{\transCom}(\syntStmt) := \forall \absEleTop \in \absEleComSet, \transCom(\top, \syntStmt)(\absEleTop) =  \transTop(\syntStmt)(\absEleTop)
\)

\(
\predicateSen_{\comJoin}(\absEleCom_0,\absEleCom_1)  := \forall \absEleTop \in \absEleTopSet, (\absEleCom_0\comJoinI\absEleCom_1)(\absEleTop) = \absEleCom_0(\absEleTop) \topJoin\absEleCom_1(\absEleTop)
\)
\end{center}

The conditions for strict context sensitivity in compositional flow insensitive analysis are similar to those in compositional flow sensitive analysis. As  the process of summarization must not lose any precision,  the same challenges exist as in the case of flow sensitive analysis. However, \toolname on flow insensitive analysis can be more friendly on memory consumption.  

\subsubsection{\toolname for Flow Insensitive Analysis}

In flow insensitive analysis, a hybrid summary \(\absEleHybrid\) is defined as a pair consisting of an abstract summary \(\absEleCom\in \absEleComSet \) and  a set of critical statements \(\syntStmts \in 2^\syntStmtset \) to be propagated:

\begin{center}
\(
\absEleHybrid  := (\absEleCom, \syntStmts) 
\)
\end{center}

Unlike in the case of flow sensitive analysis, a hybrid summary in flow insensitive analysis  does not keep the program structure. We define the join operator \(\hybridJoin\) for hybrid summaries as follows:
 
 \begin{center}
\((\absEleCom_0,\syntStmts_0)\hybridJoin (\absEleCom_1,\syntStmts_1) := 
(\absEleCom_0 \comJoinI \absEleCom_1, \syntStmts_0 \cup \syntStmts_1)
\)
\end{center}
 
 The join operator \(\hybridJoin\) just joins the two hybrid summaries component-wise.

 Hybrid summaries are computed with the following summarization function:

\begin{center}
    \(
    \begin{array}{rcl}
      \sumHybrid(\syntStmt) &  := &
    \left\{
\begin{aligned}
(\transCom(\syntStmt),\emptyset), \qquad& if \; \predicateSen_{\transCom}(\syntStmt) \\
(\bot, \{\syntStmt\}), \qquad & otherwise.
\end{aligned}
\right. \\
      \sumHybrid(\syntProg) & := &\bigsqcup^{h}_{\syntStmt\in \progToSet(\syntProg)} \sumHybrid(\syntStmt)
    \end{array}
    \)
\end{center}

The summarization function \(\sumHybrid\) just summarizes non-critical statements and keeps the critical statements in the hybrid summary. For a program \(\syntProg\), it just keeps applying the join operator \(\hybridJoin\) to the summarization of each statement in \(\progToSet(\syntProg)\). Here, we do not delay the application \(\hybridJoin\) even if \(\predicateSen_{\comJoin}\) does not hold. This may lead to precision loss, which trades precision for performance.

We define  \(\apply\) to compute the postcondition  of a hybrid summary for any precondition:
\begin{center}
    \(
    \apply((\absEleCom, \syntStmts), \absEleTop)   :=  \fixpoint_{\absEleTop}(\lambda \absEleTop'.( \absEleCom(\absEleTop')  \topJoin \bigsqcup^{t}_{\syntStmt \in \syntStmts} \transTop(\absEleTop')) ) 
    \)
\end{center}

The abstract semantics of hybrid inlining on flow insensitive analysis is defined as follows:

\begin{center}
\(
\absSemHybrid{\syntProg}(\absEleTop) := \apply(\sumHybrid(\syntProg),\absEleTop)
\)
\end{center}

\begin{theorem}
The abstract semantics of hybrid inlining on flow insensitive analysis is sound.
\end{theorem}

\begin{theorem}
The abstract semantics of hybrid inlining on flow insensitive analysis is strictly context sensitive if \(\predicateSen_{\comJoin}(\absEleCom_0,\absEleCom_1)\) holds for any \(\absEleCom_0,\absEleCom_1 \in \absEleComSet\)
\end{theorem}

\subsubsection{Context Ready Predicates} As  in the case of flow sensitive analysis, it is not necessary to 
propagate critical statements to the root procedures. The context for a hybrid summary \((\absEleCom, \syntStmts)\) is \(\absEleTop = \apply((\absEleCom,\syntStmts), \absSemTop{\syntProg}(\bot))\), where \(\syntProg\) represents the program that is not summarized in \((\absEleCom, \syntStmts)\). This context represents all possible preconditions for the given hybrid summary \((\absEleCom, \syntStmts)\). It is more precise than \(\top\). Thus, if we summarize the critical statements in \(\syntStmts\) with \(\transCom(\apply( \syntStmt, (\absEleCom,\syntStmts), \absSemTop{\syntProg}(\bot)))\), we can gain some context sensitivity compared to conventional compositional analysis.
 We define a context ready predicate  \(\predcrit(\syntStmt,\absEleTop)\), which takes as input a statement and the current context  and produces as output \(true\) if the current context is enough for the underlying specific analysis. 

 Let \(\absEleTop_c = \apply((\absEleCom,\syntStmts), \absSemTop{\syntProg}(\bot))\). Then we can
 define the following \(\contextReady\) operator  to summarize eagerly  the critical statements as follows:

\begin{center}
    \(
    \contextReady(\absEleCom, \syntStmts) := (\absEleCom \comJoinI \bigsqcup^{c}\{ \transCom(\absEleTop_c,\syntStmt) \mid 
    \syntStmt\in \syntStmts_{\mid \predcrit(\absEleTop_c)}\},\syntStmts_{\mid \neg\predcrit(\absEleTop_c)})
    \)
\end{center}
where \(\syntStmts_{\mid \predcrit(\absEleTop_c)} = \{  \syntStmt \in \syntStmts \mid \predcrit(\syntStmt, \absEleTop_c) = true\}\) represents all statements on which the current context is enough. The remaining statements are represented by \(\syntStmts_{\mid \neg\predcrit(\absEleTop_c)}\). 

With \(\contextReady\), the summarization of an invocation statement can be redefined as:

\begin{center}
    \(
      \sumHybrid(\syntax{call proc})  := \contextReady(\sumHybrid(\mapproctostmt(proc)))
    \)
\end{center}

This will allow us to summarize the critical statements wherever their accumulated context is enough. We can also define an  adequate context predicate as:

\begin{center}
    \(
\textbf{if } \predcrit(\absEleTop) =true \textbf{ then }
\forall \absEleTop' \in \absEleTopSet, \gamma(\absEleTop')\subseteq \gamma(\absEleTop) \implies
\transCom(\absEleTop, \syntStmt)(\absEleTop') =  \transTop(\syntStmt)(\absEleTop')
    \)
\end{center}

\begin{theorem}
Hybrid inlining with adequate context predicates  is sound and as precise as a version of hybrid inlining that propagates all critical statements to the root procedures.
\end{theorem}

The exact context \(\absEleTop_c = \apply((\absEleCom,\syntStmts), \absSemTop{\syntProg}(\bot))\) may be hard to know during the analysis. However, for a specific analysis given, some other information can help decide the evaluation of \(\predcrit\). For instance, recall the example discussed in Section~\ref{sec:overview}. Without any knowledge of the exact context, our analysis can still decide that the context is adequate because the receiver \prog{tx} and all its aliased variables are not accessible outside the scope of \prog{bar}1().

\subsubsection{Application Scope}

Hybrid Inlining can be applied in any flow insensitive analysis that falls into the category defined in Section~\ref{sec:flow-insen}, which is very general. 

Compared to hybrid inlining on flow sensitive analysis, hybrid inlining on flow insensitive has one advantage: the representation \((\absEleTop, \syntStmts)\) is much more concise, with only one summary and a set of statements. This alleviates the memory occupation concern significantly.

\presec
\section{Instantiation on A Pointer Analysis} \label{sec:instantiation}
\postsec

We instantiate our framework on pointer analysis, which is fundamental for many client analyses. However, compared to its top-down counterpart, compositional pointer analysis is much less explored in the literature. Thus, we design a compositional and flow insensitive pointer analysis first and then show how to apply \toolname on it. The pointer analysis follows conventional Andersen's style, but with a special constraint solving process so as to be compositional. 

\subsection{Constraint Generation}
\label{sec:pointer}

\emph{A simple language}. We characterize the pointer analysis on the following language:

\begin{center}
\(
\begin{array}{lclll}
\syntax{proc} & := & \syntStmt^{+}& \\
\syntStmt & := & \syntlv = \syntexpr \mid \syntax{call\;proc}(\syntlv_{0-n})& \\
\syntlv &:= & v \mid \var{\syntlv}.\field{f}  \mid \var{\syntlv}[c]  \mid  \var{\syntlv}[v]  & \field{f} \in \fieldset \quad v \in \allvarset \\
\syntexpr & := & \syntlv \mid c  \mid l

&  c \in \constset \quad l \in \syntlocset \\
\end{array}
\)
\end{center}

A procedure \(\syntax{proc}\) is defined as a sequence of assignment and procedure invocation statements. A left value \(\var{\syntlv}\)  can be a variable \(v\in \allvarset\), a field access \(\var{\syntlv}.\field{f}\) 
(where \( \fieldset \ni \field{f}\) denotes the set of  fields) or an index access (\ie, \(\var{\syntlv}[c]\) or \(\var{\syntlv}[v]\)). An  index can be a constant \(c\in \constset\) (e.g., string or integer) or  a variable, allowing us to  model arrays and map.  A right value expression can be a left value, a constant or a location (i.e., an allocation site \(l \in \syntlocset\)). We use \(\absloc \in  \allvarset \times  (\fieldset \cup \constset)^{*} \) to denote an access path. Concrete states are defined as \(\concEle \in \concSet :=  \allvarset  \times (\fieldset \cup \constset)^{*} \Rightarrow \constset \cup \syntlocset \), where each access path is mapped to a constant or a location. 


\begin{figure}
\begin{tabular}{l}
    \(
\begin{array}{llclclll}
\multicolumn{5}{l}{\text{Evaluation} \quad \penv: \syntlv \Rightarrow \mathcal{P}( \allvarset \times (\fieldset\cup \constset)^{*})}\\
(1)&\penv(v) (\pele) & := &\{v\}\\
(2)& \penv(\syntlv.\field{f}) (\pele) &:= &\{\absloc.\field{f}  \mid \absloc 
 \in \penv(\syntlv)\} \\
(3)& \penv(\syntlv[\const]) (\pele) &:= &\{\absloc.\const \,  \mid \absloc \in \penv(\syntlv)\}\\
(4)& \penv(\syntlv[v]) (\pele) &:=& \{\absloc.\const \,\mid  \absloc \in \penv(\syntlv),\const\in \pointsto(v,\pele)\} & \text{if} \qquad \pointsto(v,\pele) \subseteq \constset \\
(5)& \penv(\syntlv[v]) (\pele)&:= &\{\absloc.\nonconst \mid \absloc \in \penv(\syntlv)\} & \text{if} \qquad \pointsto(v,\pele) \nsubseteq \constset \\
 &&\\
 \end{array}
 \)
 \\
 \(
  \begin{array}{llcll}
\multicolumn{5}{l}{\text{Constraint Generation} \quad \consGenP: (\peleset \times \syntStmtset) \Rightarrow \peleset}\\
(6)&  \consGenP(\pele, \syntlv_0 = \syntlv_1)  &:= &  \{\absloc_0 \supseteq \absloc_1  \mid \absloc_i \in \penv(\syntlv_i,\pele), i\in \{0,1\}\} \\
(7)&  \consGenP(\pele,\syntlv = \const)  &:=&  \{\absloc \supseteq \{\const\} \mid \absloc \in \penv(\syntlv,\pele)\} \\
(8)&  \consGenP(\pele,\syntlv = \syntloc)  &:=& \{\absloc \supseteq \{\syntloc\}\mid \absloc \in \penv(\syntlv,\pele)\}  \\
(9)&    \consGenP(\pele,\syntStmts)  &:=& \bigcup_{\syntStmt \in \syntStmts}\consGenP(\pele,\syntStmt) \\
(10)&    \consGenP(\pele,\syntax{call\;proc}(\syntlv_{0-n}))  &:=& \bigcup_{\syntax{proc'}\in \pdispatch(\pele, \syntax{proc})} \consGenP(\pele, \mapproctostmt(\syntax{proc'})) \\
&&\\
\end{array}
\)
\\
\(
  \begin{array}{llcll}
   \multicolumn{5}{l}{\text{Transfer Functions} \quad \pointerTrans: (\peleset \times \syntStmtset) \Rightarrow (\peleset \Rightarrow \peleset)}\\
(11)&  \pointerTrans(\pele, \syntStmt)  &:=&  {\pele}^{'} \mapsto {\pele}^{'} \pjoin \consGenP(\pele,\syntStmt) \\
\end{array}
\)
\end{tabular}

    \caption{Abstract semantics of a pointer analysis.}
    \label{fig:pointerSemantics}
\end{figure}

\emph{Abstract states}.  An abstract state \(\pele \in \peleset\)  is a set of set constraints on access paths of the form:
 \begin{center}
\(
 \absloc \supseteq \absloc
 \quad \quad \absloc \supseteq \{l\}
 \quad \quad \absloc \supseteq \{c\}
\)
 \end{center}

This definition of \(\pele\) is classic for Anderson-style pointer analysis.  The definition of  concretization function \(\gamapointer:\peleset \Rightarrow \concSet\) is also classic and omitted. 
We define the join operator  \(\pjoin:\peleset\times\peleset \Rightarrow \peleset \) as the union of the sets of constraints: \( \pele_0\pjoin \pele_1  := \pele_0 \cup \pele_1 \).

\emph{Abstract transfer functions}. As shown in Figure~\ref{fig:pointerSemantics}, the transfer functions are defined with the help of a constraint generation function \(\consGenP: \peleset \times \syntStmtset \Rightarrow \peleset\) which transforms a statement into a set of constraints with the given context. 

For an assignment statement, \(\consGenP\) just maps the left/right values to access paths (via \(\penv\) ) and generates set relations (i.e., \(\supseteq\)) on them.  The definition of  \(\penv\) is straightforward, except for the case of \(\penv(\syntlv[v]) (\pele)\), where a variable is used as an index.  In this case, if the points-to set \(\pointsto(v,\pele)\) of \(v\) only contains constants, then each constant in \(\pointsto(v,\pele)\) is taken as an index. Otherwise, a special symbol \(\nonconst\) is introduced to represent an undecidable index. Note that the points-to set \(\pointsto(v,\pele)\) in our analysis may contain access paths involving non-local variables, which will be illustrated in Section~\ref{sec:consSolving}. The definition of \(\consGenP\) is also lifted to deal with sets of statements \(\syntStmts\), as in the Definition (9) of Figure~\ref{fig:pointerSemantics}.

For a procedure invocation statement \(\syntax{call\;proc}(\syntlv_{0-n})\), \(\consGenP\) first finds all implementations of \(\syntax{proc}\) via dispatching (denoted as \(\pdispatch(\pele,\syntax{proc})\)). Then for each implementation \(\syntax{proc'}\), \(\consGenP\) collect constraints from  all the statements \(\mapproctostmt(\syntax{proc'})\).


An abstract summary \(\pCom: \peleset \Rightarrow \peleset\) is defined as \(\pele \mapsto \pele \cup \pele_n\). That is, for each precondition \(\pele\), a summary just  adds new set constraints in \(\pele_n\) (which are generated from statements). An abstract transfer function transforms a statement into a summary, as shown in the Definition (11) in Figure~\ref{fig:pointerSemantics}. The join operator   $\comJoinP$  on summaries is defined as:
\begin{center}
    \(
    ( \pele \mapsto \pele \pjoin \pele_0) \comJoinP ( \pele \mapsto \pele \pjoin \pele_1)  :=  \pele \mapsto \pele \pjoin (\pele_0\pjoin\pele_1)
    \)
\end{center}
The other operators of this compositional pointer analysis follow the principles defined in Section~\ref{compFI}.


\emph{Critical Statements}. From the Definitions (4), (5) and (10) in Figure~\ref{fig:pointerSemantics}, we can easily identify the \emph{critical} statements: assignment statements that involve \(\syntax{lv}[v]\) and virtual procedure invocation statements (i.e., \(\syntax{call\;proc}\) where \(|\pdispatch(\top,\syntax{proc})|>1\)). On both kinds of statements, the context \(\pele\) in \(\pointerTrans(\pele, \syntStmt)\) can make a difference on the generated constraints. This compositional pointer analysis is not strictly context sensitive, because it always assumes \(top\) when summarizing these critical statements. For instance, in the Definition (10), \( \consGenP(\pele,\syntax{call\;proc}(\syntlv_{0-n})) \) would collect  constraints from all possible implementations of the invoked virtual procedure when \(\pele = \top\). This incurs precision loss. An alternative approach is to  generate disjunction here  in order to be strictly context sensitive. However, this brings significant performance loss. Furthermore, it is just impossible to generate disjunction on some critical statements, like those involving \(\syntax{lv}[v]\), since  the points-to set of the variable \(v\) is unbounded when the context is \(\top\).

\subsection{Constraints Solving}
\label{sec:consSolving}

\newsavebox{\pointerinsa} 
\begin{lrbox}{\pointerinsa}
\begin{minipage}{.22\textwidth}
\footnotesize
\begin{verbatim}
1  foo(u, v, x){
2     v.f = x;
3     return u.f;            
4  }  
\end{verbatim}
\end{minipage}
\end{lrbox}%

\newsavebox{\pointerinsb} 
\begin{lrbox}{\pointerinsb}
\begin{minipage}{.22\textwidth}
\footnotesize
\begin{verbatim}

5  getNew(){
6      return new;
7   }

\end{verbatim}
\end{minipage}
\end{lrbox}%

\newsavebox{\pointerinsc} 
\begin{lrbox}{\pointerinsc}
\begin{minipage}{.28\textwidth}
\footnotesize
\begin{verbatim}
8  bar1(x){  
9    u = getNew();
10   v = u;
11   return foo(u,v,x);
12  }  
\end{verbatim}
\end{minipage}
\end{lrbox}%

\newsavebox{\pointerinsd} 
\begin{lrbox}{\pointerinsd}
\begin{minipage}{.2\textwidth}
\footnotesize
\begin{verbatim}
13 bar2(x){  
14   u = getNew();
15   v = getNew();
16   return foo(u,v,x);
17 }  
\end{verbatim}
\end{minipage}
\end{lrbox}%

\newsavebox{\pointerinstwo} 
\begin{lrbox}{\pointerinstwo}
\begin{minipage}{\textwidth}
\footnotesize

\end{minipage}
\end{lrbox}%

\begin{figure}[t]
    \centering
\subfloat{\raisebox{0cm}{\usebox{\pointerinsa}}}
\subfloat{\raisebox{0cm}{\usebox{\pointerinsb}}}
\subfloat{\raisebox{0cm}{\usebox{\pointerinsc}}}
\subfloat{\raisebox{0cm}{\usebox{\pointerinsd}}}
    \caption{An illustrating example for constraints solving.}
    \label{fig:pintercleanCode}
\end{figure}

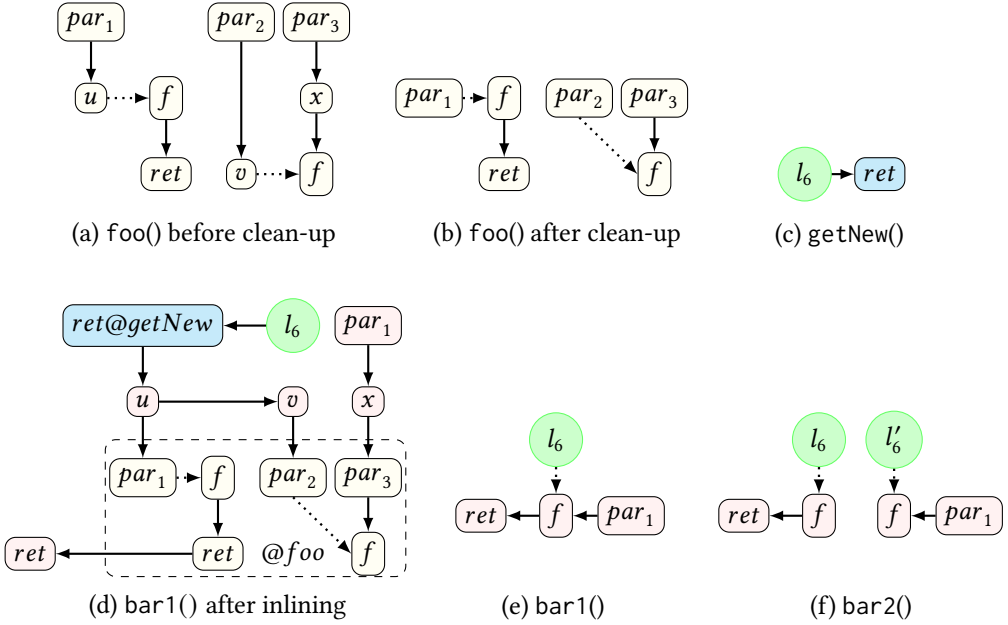
\begin{figure}[t]
\begin{tabular}{lll}
\quad\quad\scalebox{1}{  \begin{tikzpicture}[
roundnode/.style={circle, draw=green!60, fill=green!20},
squarednode/.style={rectangle, draw=black, fill=yellow!5},
squarednodegetnew/.style={rectangle, draw=black, fill=cyan!20},
]
  \begin{scope}[shift={(0, 0)}]

\node[squarednode, rounded corners]   at  (2,4)    (par2)                              {\(\var{par}_{2}\)};
\node[squarednode, rounded corners]   at  (3,4)    (par3)                              {\(\var{par}_{3}\)};
\node[squarednode, rounded corners]   at  (0,4)    (par1)                              {\(\var{par}_{1}\)};
\node[squarednode, rounded corners]   at  (0,3)    (u)                              {\(\var{u}\)};
\node[squarednode, rounded corners]   at  (2,2)    (v)                              {\(\var{v}\)};
\node[squarednode, rounded corners]   at  (3,3)    (x)                              {\(\var{x}\)};
\node[squarednode,rounded corners]     at  (1,3)      (uf)                            {\(\field{f}\)};
\node[squarednode,rounded corners]   at  (3,2)      (vf)                            {\(\field{f}\)};
\node[squarednode,rounded corners]     at  (1,2)      (ret)                            {\(\var{ret}\)};
\draw[thick, -latex] (par1.south) -- (u.north);
\draw[thick, -latex] (par2.south) -- (v.north);
\draw[thick, -latex] (par3.south) -- (x.north);
\draw[thick, -latex] (uf.south) -- (ret.north);
\draw[thick, -latex] (x.south) -- (vf.north);
\draw[line width= 0.3mm, -latex,dotted] (u.east) -- (uf.west);
\draw[line width= 0.3mm, -latex,dotted] (v.east) -- (vf.west);

\tkzputform{1.5,1.2}{\text{(a)  \prog{foo}() before clean-up}}

  \end{scope}
  
    \begin{scope}[shift={(4.5, 0)}]

\node[squarednode, rounded corners]   at  (0,3)    (u)                              {\(\var{par}_{1}\)};
\node[squarednode, rounded corners]   at  (2,3)    (v)                              {\(\var{par}_{2}\)};
\node[squarednode, rounded corners]   at  (3,3)    (x)                              {\(\var{par}_{3}\)};
\node[squarednode,rounded corners]     at  (1,3)      (uf)                            {\(\field{f}\)};
\node[squarednode,rounded corners]   at  (3,2)      (vf)                            {\(\field{f}\)};
\node[squarednode,rounded corners]     at  (1,2)      (ret)                            {\(\var{ret}\)};
\draw[thick, -latex] (uf.south) -- (ret.north);
\draw[thick, -latex] (x.south) -- (vf.north);
\draw[line width= 0.3mm, -latex,dotted] (u.east) -- (uf.west);
\draw[line width= 0.3mm, -latex,dotted] (v.south) -- (vf.west);
\tkzputform{1.7,1.2}{\text{(b)  \prog{foo}() after clean-up}}
  \end{scope}
  
    \begin{scope}[shift={(8.5, 0)}]

\node[roundnode, rounded corners]   at  (1,2)    (loc)                              {\(\var{l}_{6}\)};
\node[squarednodegetnew, rounded corners]   at  (2,2)    (ret)                              {\(\var{ret}\)};
\draw[thick, -latex] (loc.east) -- (ret.west);
\tkzputform{1.5,1.2}{\text{(c)  \prog{getNew}() }}
  \end{scope}


\end{tikzpicture}}\\
\\
\scalebox{1}{  \begin{tikzpicture}[
roundnode/.style={circle, draw=green!60, fill=green!20},
squarednode/.style={rectangle, draw=black, fill=yellow!5},
squarednodebar/.style={rectangle, draw=black, fill=red!5},
squarednodegetnew/.style={rectangle, draw=black, fill=cyan!20},
]
  \begin{scope}[shift={(0, 0)}]

\node[squarednodegetnew, rounded corners]   at  (0,5)    (newr)                              {\(\var{ret@getNew}\)};
\node[squarednodebar, rounded corners]   at  (0,4)    (baru)                              {\(\var{u}\)};
\node[squarednodebar, rounded corners]   at  (2,4)    (barv)                              {\(\var{v}\)};
\node[squarednodebar, rounded corners]   at  (3,4)    (barx)                              {\(\var{x}\)};
\node[squarednodebar, rounded corners]   at  (3,5)    (barpar)                              {\(\var{par}_1\)};
\node[roundnode, rounded corners]   at  (2,5)    (loc)                              {\(\var{l}_{6}\)};

\node[squarednode, rounded corners]   at  (0,3)    (u)                              {\(\var{par}_{1}\)};
\node[squarednode, rounded corners]   at  (2,3)    (v)                              {\(\var{par}_{2}\)};
\node[squarednode, rounded corners]   at  (3,3)    (x)                              {\(\var{par}_{3}\)};
\node[squarednode,rounded corners]     at  (1,3)      (uf)                            {\(\field{f}\)};
\node[squarednode,rounded corners]   at  (3,2)      (vf)                            {\(\field{f}\)};
\node[squarednode,rounded corners]     at  (1,2)      (ret)                            {\(\var{ret}\)};
\node[squarednodebar,rounded corners]     at  (-1.5,2)      (barret)                            {\(\var{ret}\)};
\draw[thick, -latex] (loc.west) -- (newr.east);
\draw[thick, -latex] (uf.south) -- (ret.north);
\draw[thick, -latex] (x.south) -- (vf.north);
\draw[line width= 0.3mm, -latex,dotted] (u.east) -- (uf.west);
\draw[line width= 0.3mm, -latex,dotted] (v.south) -- (vf.west);
\draw[thick, -latex] (newr.south) -- (baru.north);
\draw[thick, -latex] (baru.east) -- (barv.west);
\draw[thick, -latex] (barpar.south) -- (barx.north);
\draw[thick, -latex] (barx.south) -- (x.north);
\draw[thick, -latex] (baru.south) -- (u.north);
\draw[thick, -latex] (barv.south) -- (v.north);
\draw[thick, -latex] (ret.west) -- (barret.east);
\draw[rounded corners,dashed] (-0.5,1.7) rectangle (3.5,3.5);
\tkzputform{2,2}{@\var{foo}}
\tkzputform{1,1.3}{\text{(d)}  \; \prog{bar1}() \text{ after inlining}}
  \end{scope}

    \begin{scope}[shift={(3.5, -0.5)}]


\node[roundnode, rounded corners]   at  (2,4)    (loc2)                              {\(\var{l}_{6}\)};

\node[squarednodebar, rounded corners]   at  (3,3)    (barpar)                              {\(\var{par}_1\)};

\node[squarednodebar,rounded corners]     at  (2,3)      (uf)                            {\(\field{f}\)};

\node[squarednodebar,rounded corners]     at  (1,3)      (ret)                            {\(\var{ret}\)};

\draw[thick, -latex] (barpar.west) -- (uf.east);

\draw[thick, -latex] (uf.west) -- (ret.east);
\draw[line width= 0.3mm, -latex,dotted] (loc2.south) -- (uf.north);

\tkzputform{2,1.8}{\text{(e) \prog{bar1}() }}
  \end{scope}

    \begin{scope}[shift={(7, -0.5)}]


\node[roundnode, rounded corners]   at  (2,4)    (loc2)                              {\(\var{l}_{6}\)};
\node[roundnode, rounded corners]   at  (3,4)    (loc3)                              {\(\var{l}_{6}'\)};

 \node[squarednodebar, rounded corners]   at  (4,3)    (barpar)                              {\(\var{par}_1\)};

\node[squarednodebar,rounded corners]     at  (2,3)      (uf)                            {\(\field{f}\)};
\node[squarednodebar,rounded corners]     at  (3,3)      (vf)                            {\(\field{f}\)};

\node[squarednodebar,rounded corners]     at  (1,3)      (ret)                            {\(\var{ret}\)};

 \draw[thick, -latex] (barpar.west) -- (vf.east);

\draw[thick, -latex] (uf.west) -- (ret.east);
\draw[line width= 0.3mm, -latex,dotted] (loc2.south) -- (uf.north);
\draw[line width= 0.3mm, -latex,dotted] (loc3.south) -- (vf.north);
\tkzputform{2.6,1.8}{\text{(f)  \prog{bar2}() }}




  \end{scope}




\end{tikzpicture}}\\
\end{tabular}
    \caption{Constraints generated for the code in Figure~\ref{fig:pintercleanCode}}
    \label{fig:pointerclean}
\end{figure}

This compositional pointer analysis is a kind of partial program analysis. That means,  when a procedure is summarized, the points-to set of some variables (e.g., the parameters) in the procedure can be unknown. Thus, the constraints solving process in our analysis is different with that in conventional top-down analysis in the sense that, not only  allocation sites, but also non-local variables (e.g., the parameters)  are propagated along the constraints.  In this case, the points-to set \(\pointsto(v,\pele)\)  of a variable \(v\) can include allocation sites and non-local variables. We illustrate this with an example.

\begin{example}
\label{ex:instPointer}
In Figure~\ref{fig:pointerclean}, we show a program segment which is simple yet challenging. A pointer analysis needs to be context- and field-sensitive in order to find out that the return value of the procedure \(\prog{bar}1()\) (resp. \(\prog{bar}2()\)) is aliased (resp. not aliased) to its parameter \(\prog{x}\). Our analysis first generates constraints for the procedure \(\prog{foo}()\) in Figure~\ref{fig:pointerclean}(a). In the figure, a dotted line represents offsets. For instance, the dotted line from \(\prog{u}\) to \(\field{f}\) represents \(\prog{u}.\field{f}\).
After the propagation of the non-local variables (i.e., \(\prog{par}_{1-3}\) and \(\prog{ret}\)), we get the points-to set of \(\prog{ret}\) as \(\{\prog{par}_1.\field{f}\}\), which results in a new set relation \(ret \supseteq par_1.f\) in the summary. Similarly, our analysis gets the relation \(par_2.f \supseteq par_3\). After cleaning up local variables \(\prog{u},\prog{v}\) and \(\prog{x}\), our analysis gets the constraints in Figure~\ref{fig:pointerclean}(b). The procedure \(\prog{getNew}()\) just allocates a new memory block, as captured by the constraints in Figure~\ref{fig:pointerclean}(c) (We omit the initialization of the field \(\field{f}\) for simplicity). For the procedure \(\prog{bar}1()\), our analysis generates the constraints in Figure~\ref{fig:pointerclean}(d). After the constraints solving process, our analysis gets Figure~\ref{fig:pointerclean}(e).  Following the same principle, the summary of \(\prog{bar}2()\) is computed in Figure~\ref{fig:pointerclean}(f). Note that, there are two distinct nodes for the allocation sites at line 6 in Figure~\ref{fig:pointerclean}(f). Because the summary of \prog{getNew}() is copied at each call-site for the sake of context sensitivity. This example shows that, even though being not strictly context sensitive, our compositional pointer analysis is  context sensitive on allocation sites (\(par_1\) and \(ret\) in \(\prog{bar}2()\) would be reported aliased otherwise). 
\end{example}

While context sensitivity on allocation sites brings precision, it also incurs performance loss. The number of context sensitive allocation sites can grow rapidly in  summaries. However, it is often safe to remove them to gain performance without losing precision. For instance, in the summary of \prog{bar}1(), the access path \(l_6.f\) can be safely cleaned up, which produces \(ret \supseteq par_1\). Actually, given an access path (e.g., \(l_6.f\)) starting with an allocation site (e.g., \(l_6\)),
if its parents (e.g., \(l_6.f\))  are aliased to no variable in the summary of a procedure, then this access path can also be taken as a local variable, and safely cleaned up. Because, this access path can never be directly accessed outside the procedure. Following this principle, the access paths \(l_6.f\) and \(l_6'.f\) in \prog{bar}2()  can also be cleaned up. Our implementation does this optimization for the sake of performance.

Even though this pointer analysis is context sensitive to some extent, it loses context sensitivity on two kinds of critical statements: assignment statements that involve \(\syntax{lv}[v]\) and virtual procedure invocation statements. 
  \toolname can make it strictly context sensitive on these two kinds of critical statements.



\subsection{\toolname on This Pointer Analysis}
\label{sec:inst:hi}

To apply \toolname, following the framework in Section~\ref{sec:flow-insen}, we just need to define  \(\predicateSen_{\pointerTrans}\) (to identify the critical statements). In this pointer analysis, the two kinds of critical statements can be easily identified with syntactic pattern matching and class hierarchy.

Defining \(\predicateSen_{\pointerTrans}\)  is enough to apply \(k\)-limit Hybrid Inlining, (i.e., propagating every critical statement \(k\) steps). However, we can also define  the context ready predicate \(\predcrit\) to allow  critical statements to be propagated the steps as they need. We define an adequate context predicate \(\predcrit_{a}\) for statements involving index access (\ie \(\syntlv[v]\)) :

\[
\predcrit_{a}(\syntlv[v]=\syntexpr, \pele) := \pointsto(v,\pele) \cap \freevariables(\pele) = \emptyset 
\]
where \(\freevariables(\pele)\) denotes the set of  variables that are accessible outside the current procedure. This predicate indicates that the propagation stops when the points-to set of the index variable \(v\) does not contain any free variables, i.e.,  it cannot change any more. Thus, this context is adequate. Note that the variables in the critical statements in the hybrid summary are also free variables. 

We also need an adequate context predicate \(\predcrit_{a}\) for virtual invocation statements (\ie \(\syntlv = \syntax{proc}(\syntlv_{0-n})\)), which is defined as:

\[
\hspace{-3ex}
    \begin{aligned}
\predcrit_{a}(\syntlv = \syntax{proc}(\syntlv_{0-n}),\pele) :=
\pointsto(\syntlv_0,\pele) \cap \freevariables(\pele) = \emptyset \\
\vee | \pdispatch(\syntax{proc}, \pele)| = 1
   \end{aligned}
\]
        
This predicate checks whether the points-to set of the receiver \(\syntlv_0\) contains free variables. It also checks whether the callsite is monomorphic, which can happen when the type information gathered on the receiver has indicated such a unique implementation. This condition is adequate if the analysis does not consider error states (when, \eg, the receiver points to \(null\)).


\newsavebox{\insthybirdone} 
\begin{lrbox}{\insthybirdone}
\begin{minipage}{.24\textwidth}
\footnotesize
\begin{verbatim}
1  getP(map,key){
2    return map[key];
3  }
\end{verbatim}
\end{minipage}
\end{lrbox}%

\newsavebox{\insthybirdtwo} 
\begin{lrbox}{\insthybirdtwo}
\begin{minipage}{.24\textwidth}
\footnotesize
\begin{verbatim}
4  setP(map,key,v){
5    map[key]=v; 
6  }
\end{verbatim}
\end{minipage}
\end{lrbox}%

\newsavebox{\insthybirdthree} 
\begin{lrbox}{\insthybirdthree}
\begin{minipage}{.24\textwidth}
\footnotesize
\begin{verbatim}
7  build(u){
8   v = new;
9    setP(v,"old", getP(u,"cur"));
10  return v;}
\end{verbatim}
\end{minipage}
\end{lrbox}%

\begin{figure}
    \centering
\subfloat{\raisebox{0.5cm}{\usebox{\insthybirdone}}}
\subfloat{\raisebox{0.5cm}{\usebox{\insthybirdtwo}}}
\subfloat{\raisebox{0cm}{\usebox{\insthybirdthree}}}
    \caption{An illustrating example for Hybrid Inlining on the pointer analysis.}
    \label{fig:hypointercode}
\end{figure}

\newsavebox{\insthybridgraph} 
\begin{lrbox}{\insthybridgraph}
\begin{minipage}{.8\textwidth}
  \begin{tabular}{  c  c  cc}
         	\raisebox{0\height}{ \scalebox{1}{  \begin{tikzpicture}[
roundnode/.style={circle, draw=green!60, fill=green!5},
squarednode/.style={rectangle, draw=black, fill=yellow!5},
squaredfield/.style={rectangle,dotted,thick, draw=black, fill=yellow!5},
]
  \begin{scope}[shift={(0, 0)}]

\node[squarednode, rounded corners]   at  (2,3)    (par1)                              {\(\var{par}_{2}\)};
\node[squarednode, rounded corners]   at  (2,2)    (key)                              {\(\var{key}\)};
\node[squarednode,rounded corners]     at  (1,3)      (par0)                            {\(\var{par}_1\)};
\node[squarednode]   at  (1,2)      (ext)                            {\(\var{map}\)};
\node[squarednode,rounded corners,ultra thick]   at  (1.5,1)      (stmt)                            {\(ret = map[key]\)};
\node[squarednode,rounded corners]     at  (1,0)      (ret)                            {\(\var{ret}\)};
\draw[thick, -latex] (par1.south) -- (key.north);
\draw[thick,-latex] (par0.south) -- (ext.north);
\draw[thick, -latex] (ext.south) -- (1,1.3);
\draw[thick, -latex] (key.south) -- (2,1.3);
\draw[thick, -latex] (1,0.7) -- (ret.north);
  \end{scope}
\end{tikzpicture}}}
           &  	\raisebox{0\height}{ \scalebox{1}{  \begin{tikzpicture}[
roundnode/.style={circle, draw=green!60, fill=green!5},
squarednode/.style={rectangle, draw=black, fill=yellow!5},
buildnode/.style={rectangle, draw=black, fill=red!5},
squaredfield/.style={rectangle,dotted,thick, draw=black, fill=yellow!5},
]
  \begin{scope}[shift={(0, 0)}]

\node[buildnode,rounded corners]     at  (2,4)      (cur)                            {``\(\var{cur}\)''};
\node[buildnode,rounded corners]     at  (1,4)      (buildthis)                            {\(\var{u}\)};
\node[squarednode, rounded corners]   at  (2,3)    (par1)                              {\(\var{par}_{2}\)};
\node[squarednode, rounded corners]   at  (2,2)    (key)                              {\(\var{key}\)};
\node[squarednode,rounded corners]     at  (1,3)      (this)                            {\(\var{par}_1\)};
\node[squarednode]   at  (1,2)      (ext)                            {\(\var{map}\)};
\node[squarednode,rounded corners,ultra thick]   at  (1.5,1)      (stmt)                            {\(ret = map[key]\)};
\node[squarednode,rounded corners]     at  (1,0)      (ret)                            {\(\var{ret}\)};
\draw[rounded corners,dashed] (0.2,-0.5) rectangle (2.8,3.5);
\tkzputform{2,0}{@\var{getP}}
\draw[thick, -latex] (buildthis.south) -- (this.north);
\draw[thick, -latex] (cur.south) -- (par1.north);
\draw[thick, -latex] (par1.south) -- (key.north);
\draw[thick, -latex] (this.south) -- (ext.north);
\draw[thick, -latex] (ext.south) -- (1,1.3);
\draw[thick, -latex] (key.south) -- (2,1.3);
\draw[thick, -latex] (1,0.7) -- (ret.north);
  \end{scope}
\end{tikzpicture}}}
     &  	\raisebox{0\height}{ \scalebox{1}{  \begin{tikzpicture}[
roundnode/.style={circle, draw=green!60, fill=green!5},
squarednode/.style={rectangle, draw=black, fill=yellow!5},
buildnode/.style={rectangle, draw=black, fill=red!5},
squaredfield/.style={rectangle,dotted,thick, draw=black, fill=yellow!5},
]
  \begin{scope}[shift={(0, 0)}]


\node[buildnode,rounded corners]     at  (1,4)      (buildthis)                            {\(\var{u}\)};

\node[squarednode,rounded corners]     at  (1,3)      (this)                            {\(\var{par}_1\)};
\node[squarednode]   at  (1,2)      (ext)                            {\(\var{map}\)};

\node[squarednode]   at  (1,1)      (cur)                            {\(\var{cur}\)};

\node[squarednode,rounded corners]     at  (1,0)      (ret)                            {\(\var{ret}\)};
\draw[rounded corners,dashed] (0.3,-0.5) rectangle (1.8,3.5);
\draw[thick, -latex] (buildthis.south) -- (this.north);
\draw[thick, -latex] (this.south) -- (ext.north);
\draw[thick, -latex] (cur.south) -- (ret.north);
\draw[line width= 0.5mm, -latex,dotted] (ext.south) -- (cur.north);
  \end{scope}
\end{tikzpicture}}}
&\raisebox{0\height}{ \scalebox{1}{  \begin{tikzpicture}[
roundnode/.style={circle, draw=black!60, fill=red!5, minimum size = 5pt,inner sep = 2pt},
squarednode/.style={rectangle, draw=black, fill=red!5},
squaredfield/.style={rectangle,dotted,thick, draw=black, fill=red!5},
]
  \begin{scope}[shift={(0, 0)}]

\node[squarednode,rounded corners]     at  (1,3)      (this)                           {\(\var{par}_{1}\)};
\node[squarednode]   at  (1,2)      (pi)                            {\(\var{cur}\)};
\node[squarednode,rounded corners]     at  (2,3)      (ret)                           {\(\var{ret}\)};
\node[squarednode]   at  (2,2)      (pir)                            {\(\var{old}\)};
\draw[thick, -latex] (pi.east) -- (pir.west);
\draw[line width= 0.5mm, -latex,dotted] (this.south) -- (pi.north);
\draw[line width= 0.5mm, -latex,dotted] (ret.south) -- (pir.north);
  \end{scope}
\end{tikzpicture}}}
\\
(a) \(\mathtt{getP}()\) & (b) a part of \(\mathtt{build}()\) & (c) a part of  \(\mathtt{build}()\)  & (d) \(\mathtt{build}\)()\\
  \end{tabular}
\end{minipage}
\end{lrbox}%

\begin{figure}
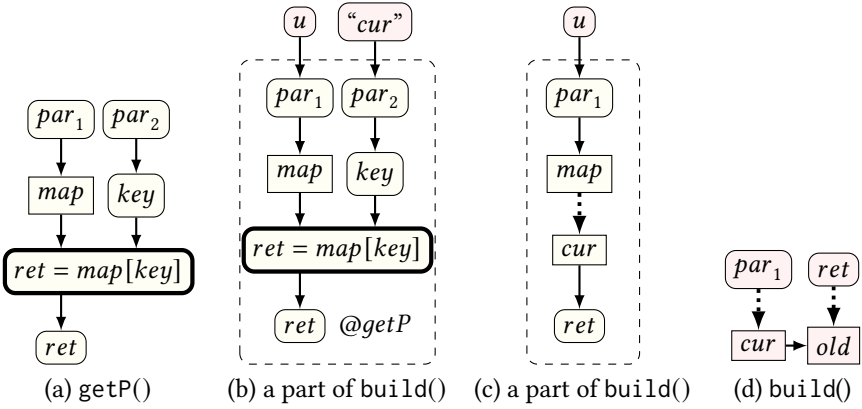

    \centering
\usebox{\insthybridgraph}
    \caption{Constraints generated for the code in Figure~\ref{fig:hypointercode}.}
    \label{fig:hypointercons}
\end{figure}

\begin{example}
Figure~\ref{fig:hypointercode} shows a code segment, where two \emph{map} accesses at line 2 and line 5 are wrapped by two procedures \(\prog{getP}()\) and \(\prog{setP}()\).  Without \toolname, our pointer analysis would just totally lose context- and index-sensitivity on these map accesses, because the value of the index (stored in the variable \(\prog{key}\)) is unknown when  \(\prog{getP}()\) and \(\prog{setP}()\) are summarized.  Note that, it is impossible to create disjunction here, because the possible values stored in \(\prog{key}\) are unbounded.

With \toolname, our analysis identifies  \prog{ret@getP=map[key]} as critical (by syntactic checking), and decides to propagate it, because the points-to set of \prog{key} includes the free variable \(\mathtt{par}_2@\mathtt{getP}\) as showed in Figure~\ref{fig:hypointercons}(a).  Inlining this summary to the procedure \(\prog{build}()\), our analysis gets  Figure~\ref{fig:hypointercons}(b) (only some constraints are shown for the sake of concision).
After cleaning \(\var{par}_2@\var{getP}\), the points-to set of \prog{key}  only contains the constant ``\( \var{cur}\)''. Thus
the critical statement is summarized, producing Figure~\ref{fig:hypointercons}(c).  
 With a similar process  applied to   \prog{setP}(),  we obtain the summary of \prog{build}() as shown in Figure~\ref{fig:hypointercons}(d).  This  summary achieves index- and context- sensitivity, thanks to \toolname. 

\end{example}



\presec
\section{Evaluation} \label{sec:evaluaiton}
\postsec

We have implemented our compositional pointer analysis in Java based on GraalVM~\cite{graalvm}, which transforms Java bytecode to an SSA-like graph representation. That alleviates the impression brought by flow insensitivity of our pointer analysis.  It is very simple to apply \toolname. We just need to add some conditions for each kind of critical statements to decide whether they should be propagated or summarized. We have implemented  Hybrid Inlining with both \(k\)-limit propagation steps and adequate context predicates. We evaluate our approach by addressing the following research questions:

\begin{itemize}
    \item[\bf RQ1:] What are the benefits of applying \toolname on conventional compositional analysis with respect to precision and performance?
    \item[\bf RQ2:] Why does \toolname achieve such benefits? 
    \item[\bf RQ3:] What is the practical significance of our work when applied in the industry?
\end{itemize}

To our best knowledge, our approach is the \emph{first} work on selective context sensitivity for compositional analysis. We take some selective context sensitivity works~\cite{li2018precision,lu2021selective} for top-down analysis implemented in~\cite{he2022qilin} as a comparison. It is worth mentioning that, we do not intend to demonstrate which approach is better via this comparison, because their approaches and ours are built on very different grounds (top-down analysis vs compositional analysis).  The comparison is conducted only to help evaluate the benefits of our work.

\subsection{RQ1: Precision and Performance of \toolname}

\definecolor{ljcGreen}{RGB}{40,200,5}
\definecolor{ljcRed}{RGB}{200,40,5}
\newcommand{\tp}{{\color{ljcGreen}\checkmark}}
\newcommand{\fp}{{\color{ljcRed}\(\otimes\)}}

We first measure the precision of Hybrid Inlining on a micro-benchmark suite \emph{PointerBench}~\cite{spath2016boomerang}, which comes with ground truth. All its test cases   are listed in Table~\ref{table:pinterBench}. 
 We have also added the example from Section~\ref{sec:overview} to the benchmark suite.

 The artifact from~\cite{he2022qilin} provides a bunch of selective context sensitive works for top-down pointer analysis.  We choose Zipper~\cite{li2018precision} and SELECTX~\cite{lu2021selective},  because they are relatively more recent among all supported options. As for the baseline analyses, we run the top-down context insensitive analysis (Column ``TopCI'' in Table~\ref{table:pinterBench}) and the 3-call-site sensitive analysis (Column ``3C''). We choose 3-call-site sensitivity, because that is the maximal context sensitivity needed on these micro-benchmarks. The results of Zipper/SELECTX guided selective 3-call-site sensitive analyses are shown in the Columns ``Z-3C''/``S-3C''.  We also run our compositional pointer analysis (Column ``ComCI'') and that analysis  with \(3\)-limit \toolname (Column ``HI3''), where each critical statement is propagated 3 steps.

In the experimental results from Table~\ref{table:pinterBench}, each \tp (resp. \fp ) represents a true (resp. false) positive. From the table, we can see that, HI3 is the most precise analysis.  The top-down context insensitive analysis (i.e., TopCI) is the most imprecise one. However, precision is not just a simple number. This table also reveals where each analysis gains/loses precision. In fact, the most imprecise TopCI is  more precise than HI3 on some cases (e.g., \emph{exception}), because, it models control flows involving exceptions more precisely. Our HI3 is only more precise than 3C on the test cases in the \emph{Containers} category, since our implementation models containers more precisely (as shown in Section~\ref{sec:instantiation}). Context sensitivity is actually only needed for the test cases \emph{returnValue}, \emph{contextSensitivity}, \emph{objectSensitivity} and  \emph{overviewExample}. On these cases, 3C, S-3C and HI3 lose no precision. However, Zipper loses precision on the case \emph{returnValue}, where context sensitivity is needed for the allocation site in an initialization procedure. Z-3C fails to identify that procedure as critical, thus producing a false positive.

\begin{table*}
\caption{Precision Comparison on PointerBench.}

\begin{tabular}{|c|m{2.5cm}|m{1.35cm}<{\centering}|m{1.35cm}<{\centering}|m{1.35cm}<{\centering}|m{1.35cm}<{\centering}|m{1.35cm}<{\centering}|m{1.35cm}<{\centering}|}
\hline
\multicolumn{2}{|c|}{Tests}    & TopCI  & 3C  & Z-3C   & S-3C& CompCI & HI3 \\
\hline
\multirow{7}{*}{\rotatebox{90}{Basic}}&branching & \tp\tp &\tp\tp   &  \tp\tp   &  \tp\tp   &  \tp\tp& \tp\tp\\
&interprocedural &\tp\tp &\tp\tp   &  \tp\tp   &   \tp\tp&  \tp\tp& \tp\tp\\
&loops &\tp\tp\tp\fp & \tp\tp\tp\fp&\tp\tp\tp\fp&\tp\tp\tp\fp&\tp\tp\tp\fp&\tp\tp\tp\fp\\
&parameter & \tp\tp\tp\tp& \tp\tp\tp\tp&\tp\tp \tp\tp&  \tp\tp\tp\tp& \tp\tp\tp\tp& \tp\tp\tp\tp\\
&recursion & \tp& \tp&\tp& \tp& \tp & \tp\\
&returnValue & 4 \tp,\fp &   4 \tp & 4 \tp,\fp & 4 \tp& 4 \tp &  4 \tp\\
&simpleAlias & \tp\tp &  \tp\tp & \tp\tp  & \tp\tp &\tp\tp & \tp\tp\\
\hline
\multirow{4}{*}{\rotatebox{90}{Containers}}&array & \tp\tp\fp & \tp\tp\fp & \tp\tp\fp&  \tp\tp\fp & \tp\tp& \tp\tp\\
&list & 4\tp, 2\fp& 4\tp, 2\fp & 4\tp, 2\fp &4\tp, 2\fp &4 \tp, 2\fp& 4\tp, 2\fp\\
&map & \tp\tp\fp & \tp\tp\fp & \tp\tp\fp & \tp\tp\fp & \tp\tp& \tp\tp\\
&set & \tp\fp & \tp\fp & \tp\fp & \tp\fp&  \tp\fp& \tp\fp\\
\hline
\multirow{6}{*}{\rotatebox{90}{General Java}}&accessPath & \tp\tp & \tp\tp & \tp\tp &  \tp\tp&\tp\tp&\tp\tp\\
&fieldSensitivity & \tp\tp\tp\tp &\tp\tp \tp\tp &\tp\tp \tp\tp&\tp\tp \tp\tp&\tp\tp \tp\tp&\tp\tp \tp\tp\\
&contextSensitivity & 6\tp, 3\fp & 6\tp & 6\tp  &6\tp&6\tp,3\fp&6\tp\\
&flowSensitivity &  \tp\ &\tp &\tp &\tp&\tp&\tp\\
&objectSensitivity & 4\tp, 4\fp &4\tp &4\tp  & 4\tp& 4\tp &4\tp \\
&strongUpdate &\tp\tp\tp\fp&\tp\tp\tp\fp&\tp\tp\tp\fp&\tp\tp\tp\fp&\tp\tp\tp\fp&\tp\tp\tp\fp\\
\hline
\multirow{6}{*}{\rotatebox{90}{Cornor Cases}}&exception &\tp\tp&\tp\tp&\tp\tp&\tp\tp&\tp\tp\fp&\tp\tp\fp\\
&interface &\tp\tp&\tp\tp&\tp\tp&\tp\tp&\tp\tp&\tp\tp\\
&null &&&&&&\\
&outerClass&\tp\tp\fp&\tp\tp\fp&\tp\tp\fp&\tp\tp\fp&\tp\tp\fp&\tp\tp\fp\\
&staticVariable&\tp\tp&\tp\tp&\tp\tp&\tp\tp&\tp\tp&\tp\tp\\
&superClass& \tp\tp\fp & \tp\tp\fp &  \tp\tp\fp & \tp\tp\fp& \tp\tp\fp& \tp\tp\fp \\
\hline
\multicolumn{2}{|c|}{overviewExample} & \tp\fp & \tp\tp & \tp\tp  & \tp& \tp\fp& \tp\tp\\
\hline
\multicolumn{2}{|c|}{Precision}& 79.6\% & 86.8\% & 85.5\% & 86.8\% & 83.1\% & 88.1\% \\
\hline
\end{tabular}
\label{table:pinterBench}
\end{table*}

To evaluate the  performance and precision of \toolname on large applications,  we run all Dacapo~\cite{blackburn2006dacapo} benchmarks in the artifact of \cite{he2022qilin} except \emph{jython}, because our analysis (even in a context insensitive setting) just explodes with the number of access paths on it.  All experimental results are  shown in Table~\ref{table:CFCIandCFCS}. We run the top-down context insensitive analysis (Column ``TopCI''),  the 2-call-site sensitive analysis (Column ``2C'') and Zipper/SELECTX guided selective 2-call-site sensitive analyses (Columns ``Z-2C''/``S-2C''). Our compositional pointer analysis is run with three different configurations:(1) context insensitive (Column ``ComCI''); (2)  \toolname where each critical statement is propagated 2 steps (Column ``HI2''); (3) \toolname with adequate context predicates defined in Section~\ref{sec:inst:hi} (Column ``HIA''). In the third configuration, we  have also implemented a simple heuristic to limit the propagation of some critical statements even if  adequate context predicates are not satisfied, for the sake of performance. That is,
if the number of critical statements reaches a threshold in a summary, our analysis summarizes the critical statements that propagate the most steps immediately.  This is to avoid the number of critical statement to explode in some extreme cases. All experiments are run on a personal laptop with a 2.6 GHz CPU and 16 GB memory. The time budget is set to be one hour.

\newcommand{\mReach}{\#reach-procs}
\newcommand{\mTime}{time(s)}
\newcommand{\mPrecision}{\#poly-call}

\renewcommand\arraystretch{1.2}
\begin{table*}
\caption{Evaluation of Performance and Precision on Dacapo}
\setlength{\tabcolsep}{1.34ex}
\begin{tabular}{|c|c|c|c|c|c|c|c|c|c|c|c}
\hline
Tests&  Metrics &TopCI & 2C &Z-2C & S-2C & ComCI & HI2 &HIA \\
\hline
\multirow{3}{*}{\rotatebox{90}{antlr}}    & 
 \mReach  & 8194 &  8023& 8042 &8023 &5280 & 5173 &5129 \\
          &\mPrecision & 1987 & 1852&1870 &1852 &1891 & 1786 & 1692\\
          & \mTime &22.8& 3035.8 &75.3 &356.3 &11.6 &13.5&13.4\\
         \hline
         \multirow{3}{*}{\rotatebox{90}{bloat}}    &
\mReach & 9464 &9277& 9307&9277 &8233 & 7657 &7554\\
          &\mPrecision & 2344&2089 &2119 &2089 & 2627 &2285 &2066\\  
          & \mTime &17.9 &3387.1&928.7& 2749.0 &18.9 &22.5&27.4\\
         \hline
         \multirow{3}{*}{\rotatebox{90}{chart}}    & 
 \mReach &15933 & TO &15617 & 15521 & 16187 &15835 & 15827\\        
 &\mPrecision &2732 & TO&2493& 2469 &6529 & 5939&5503\\
 & \mTime &56.5 &TO &934.9& 1136.9 &41.2 & 48.3 & 62.8\\
                  \hline
         \multirow{3}{*}{\rotatebox{90}{eclipse}}     
& \mReach & 23387 &TO&TO&TO & 23033 & 20829&19218\\
          &\mPrecision &10738&TO&TO&TO & 13535& 12583 & 9602\\
          & \mTime & 118.9&TO&TO&TO &102.1 &129.0&200.0 \\
                  \hline
         \multirow{3}{*}{\rotatebox{90}{fop}}    & 
 \mReach & 8001 &TO&7845 &7826 & 19302 & 169397 & 14461\\
          &\mPrecision & 1223&TO&1092 &1073& 7530& 5295 & 4187\\
          & \mTime & 34.9&TO&65.3&223.4&66.4 &60.1 & 49.4\\
                  \hline
         \multirow{3}{*}{\rotatebox{90}{hsqldb}}    
& \mReach &7389 &TO&7222 &7203 &7785 &7708 &7695 \\
          &\mPrecision &1213 &TO& 1087 & 1067 & 1902 & 1798 & 1718\\
                   & \mTime &  36.9&TO &49.5 &226.9 &16.0&22.8&30.5\\
                  \hline
         \multirow{3}{*}{\rotatebox{90}{luindex}}    & 
 \mReach &7413  &TO& 7259 &7241 & 7548 &7426 &7391 \\
          &\mPrecision & 1294 &TO& 1166 &1147 & 2682 &2505 &2400 \\
          & \mTime &22.6 & TO& 48.2 &236.2 & 13.3 & 16.4 &15.1\\
                  \hline
         \multirow{3}{*}{\rotatebox{90}{lusearch}}    
& \mReach & 8098 & TO&7941 &7923 & 6767 & 6659 &6646 \\
          &\mPrecision &1505 &TO& 1372& 1351 &2245 & 2061 &2011 \\
         & \mTime &30.7& TO&43.6& 239.9 & 19.1 & 18.2 &27.8\\
                  \hline
         \multirow{3}{*}{\rotatebox{90}{pmd}}    
& \mReach &12369 &TO&12208 &12192 & 10396 &9945  & 9879\\
          &\mPrecision &2989 &TO&2805 &2789 & 2806&2546 & 2459 \\
         & \mTime & 68.6 &TO&385.9 &821.0& 22.6 &42.5 &28.1 \\
                  \hline
                  \multirow{3}{*}{\rotatebox{90}{xalan}}    
& \mReach &10096 &TO &9937 &9918 &14240&12814 & 12190 \\
          &\mPrecision &2101 &TO &1939 &1918 &6781 & 6081 &5965\\
         & \mTime &46.0 & TO& 102.2 & 507.2 & 56.4&85.1 &218.9\\
\hline
\end{tabular}
\label{table:CFCIandCFCS}
\end{table*}


On each test case, we report the  number of reached procedures (Row ``\mReach''), the number of polymorphic call-sites (Row ``\mPrecision'') and running time (Row ``\mTime''). Polymorphic call-sites are used for measuring precision (smaller is better) and running time is for performance.  Table~\ref{table:CFCIandCFCS} shows that both HI2 and HIA are significantly more precise than ComCI with little performance loss. HI2 and HIA reduce the number of polymorphic call-sites by 10.4\% and 17.6\%  on average respectively. The recorded running times are all within 2X of ComCI except for the case of \emph{xalan}. On some test cases like \emph{fop}, HI2 and HIA are even faster than ComCI. On average, HI2 and HIA brings 27.0\% and 64.5\% additional overhead on running time. Compared with  HIA  and HI2, HIA requires more running time, but is significantly more precise than HI2.

As for top-down analysis, Z-2C and S-2C achieve an average of 8.7\% and  9.8\% reduction on the number of polymorphic call-sites with 10.6X and 26.3X increase on running time (TO represents timeout and is counted as 3600s)  from TopCI. It is worth mentioning that, the running times recorded in our reproduction is different with those reported in their papers~\cite{li2018precision,lu2021selective}. The reason, we suspect, is that the hardware used in their experiments is much more powerful than ours. For instance, a Xeon E5-1660 3.2GHz machine with 256GB of RAM is used in~\cite{lu2021selective}. This also influences the ratio of time increase with their approaches. In \cite{lu2021selective}, the authors report an average of 6.8X increase on running time (with the same setting as S-2C) on all test cases in Table~\ref{table:CFCIandCFCS} except for \emph{bloat} and \emph{lusearch}.  On the same test cases, our reproduction reports 13X time increase. 

While it is reasonable to evaluate different selective context sensitivity works on their respective grounds, we cannot compare the precision metrics in Table~\ref{table:CFCIandCFCS} between top-down analysis and compositional analysis directly. Because the ground truth is absent and neither of the two implementations is strictly sound. The unsoundness comes from the front-end (e.g., Soot and Wala)~\cite{prakash2021effects} , rather than the analysis algorithms, which is also noticed by~\cite{he2022qilin}. The main reason is that some libraries are not successfully loaded~\cite{prakash2021effects}, or are set to be ignored. In fact, TopCI reports 758 call-sites without call-edges on \emph{antlr}, while ComCI reports 36. However, this does not mean that our implementation is more sound (because the reached  procedures are different).

\subsection{RQ2: Why does \toolname achieve such benefits?}

\begin{table*}
\caption{Critical statements and their propagation}
\begin{tabular}{|c|m{1.1cm}<{\centering}|m{1.1cm}<{\centering}|m{1.1cm}<{\centering}|m{1.1cm}<{\centering}|m{1.1cm}<{\centering}|m{1.1cm}<{\centering}|m{1.1cm}<{\centering}|m{1.1cm}<{\centering}|m{1.1cm}<{\centering}|}
\hline
Tests & \(\#C\) & \(\#C_{prop}\) & \(\%C_{prop}\)  & \(K_{max}\) & \(K_{avg}\) &Proc-H & Proc-Z & Proc-S\\
\hline
antlr&	3854&	2224&	0.93\%	&10&	3.25	&18.5\% &	39.4\% &	72.4\%\\ 
bloat&	6443&	3353&	1.20\%	&17	&4.57&	19.9\% &	40.2\% &	71.5\% \\
chart&	19512&	15430&	2.57\%	&9&	2.52 &24.6\% &	32.6\% &	67.1\% \\
eclipse	&28350&	12341&	1.43\%&	26&	3.30	&27.2\% & -& -\\	
fop	&16433&	9293&	1.77\%&	10	&1.71	&21.3\% &	31.4\% &	66.9\% \\
hsqldb	&6651&	2574&	0.87\%	&6&	1.82&16.6\% &	35.1\% &	69.2\% \\
luindex&	6668&	3747&	1.44\%	&19	&2.81	&22.7\% &	34.4\% &	69.1\% \\
lusearch	&5909	&4166&	1.82\%	&13&	2.55 &25.1\% &	34.5\% &	69.2\% \\
pmd	&6909	&3102&	0.90\%	&10&	2.11		&19.2\% &	39.3\% &	72.4\% \\
xalan	&57055&	10594&	1.89\%&	38&	5.22& 28.7\% &	40.2\% &	70.1\% \\

\hline
\end{tabular}
\label{table:insights}
\end{table*}

To understand how \toolname achieves such benefits, we provide some insights on how critical statements are propagated. Table~\ref{table:insights} shows the number of critical statements (Column ``\(\#C\)''), the critical statements that are propagated (``\(\#C_{prop}\)''), the ratio of such critical statements in all statements (Column ``\(\%C_{prop}\)'',  counted in the level of Graal IR),  the maximal and average propagation steps (Columns  ``\(K_{max}\)'' and ``\(K_{avg}\)'') in the HIA configuration (i.e., \toolname with adequate context predicates).  From this table, we can see that, the critical statements that need to be propagated only account for a very small potion of all statements (less than 2\% in all test cases except \emph{chart}). This explains why \toolname delivers excellent performance: it only re-analyzes critical statements rather than non-critical statements for different contexts. As for the propagation steps, while \toolname allows individual critical statements to be propagated really far (up to 38), the average propagation steps are between 1.71 and 5.22. This explains why HIA does not cause too much performance loss, and why HIA is more precise than HI2.

We also report the portion of procedures that get strict context sensitivity because of the propagation of critical statements, as listed in Column the ``Proc-H''. As a comparison, we also list the portion of procedures that Zipper and SELECTX assign context sensitivity in the Columns ``Proc-Z'' and ``Proc-S''. We can see that, \toolname gives context sensitivity to fewer procedures than both Zipper and SELECTX. The reason is, the underlying compositional pointer analysis is actually context sensitive to some extent (e.g., on field accesses as shown in Section~\ref{sec:instantiation}). This also explains why 
\toolname achieves the excellent performance in our evaluation.

\subsection{RQ3: Applications in the Industry}

We also evaluate our work on 10 real industrial applications, which are   all the core modules of a large financial transaction system from a company providing worldwide financial services. These applications are all in microservice architecture, with each consisting of dozens or hundreds of entry procedures. Given the difficulty of faking a sound single main procedure for these applications to enable top-down analysis, we only run our compositional pointer analysis here (by traversing all application procedures). The experimental results are shown in Table~\ref{table:industry}.   The Columns ``Tests'' and ``Size'' denote the anonymized names and sizes (measured in terms of the number of classes and the size of bytecode) of these applications. These applications heavily use   libraries, which are usually more than ten times as large as the application code. However, to ensure termination, we set the maximal depth  as 3 in dealing with library procedures. For both conventional compositional analysis (Column ``ComCI'') and Hybrid Inlining (Column ``HIA''), we report the number of reached procedures (Column ``\#reach''), polymorphic call-sites (Column ``\#poly'') and running time (Column ``time(s)''). We also report the portion of propagated critical statements (Column ``\%prop''), the maximal and average propagation steps (Columns \(k_{avg}\) and \(k_{avg}\)). We can see that, while gaining precision, HIA is surprisingly faster than ComCI in 6 out of 10 test cases. The additional time overhead brought by HIA is only 1\% on average. There are two reasons for this: (1) the portion of  propagated critical statements is very low. This is because, these industrial applications deal with much simpler tasks (like transactions) than the  aforementioned open source applications; (2) These industrial applications use heavily  callbacks (e.g., with \prog{TransactionTemplate} in the \emph{Spring} framework). Imprecise dispatching on their virtual call-sites can lead to huge performance loss (and precision loss).



\begin{table*}
\centering
\caption{Evaluation on Industrial Applications}
\scalebox{.96}{
\addtolength{\tabcolsep}{-.3ex}
\begin{tabular}{|c|c|c|c|c|c|c|c|c|c|c|c|}
\hline
\multirow{2}{*}{Tests} & \multicolumn{2}{c}{Size} &  \multicolumn{3}{|c|}{ComCI} & \multicolumn{6}{|c|}{HIA} \\
\cline{2-12}
& \(\#class\)&  \(byteCode\) & \#reach &\#poly& time(s) & \#reach  & \#poly & time(s) & \%prop & \(k_{max}\) & \(k_{avg}\) \\
\hline
app1  & 613  & 207K & 7428	&178&	9.6 &	7414&	127&	9.2& 	0.29 & 7 &2.39 \\ 
app2  & 7339 & 1718K  &82162&	1304&	91.2 &	80728&	1233	&101.5& 	0.24& 20 & 2.31\\ 
app3  & 2174 &  576K   & 23994&	559	&26.9 &	23989&	532	&22.8 &	0.15 & 8 & 1.66\\ 
app4  & 2279 &  641K    & 25451&	937	&28.3 &	25436&	839	&34.8 &	0.21 & 8 & 1.86 \\ 
app5  & 3842 & 986K    &  24903&	1060&	24.3 &	24901&	526	&27.9 &	0.47& 14 & 2.62 \\ 
app6  & 3758  & 700K   & 38805&	468&	51.9 &	38773&	399	&51.2 &	0.10 & 8 & 1.60\\ 
app7  & 3003 & 950K  &30724&	295	&37.9 &	30705&	247&	40.1 &	0.19 & 18 & 3.01\\ 
app8  & 4375 & 1419K    & 43627&	3355&	94.0 &	43612&	2824&	85.1 &	0.38& 13 & 2.32\\ 
app9  & 1391  & 216K    & 15252&	277	&22.6 &	15289	&242	&22.4 &	0.16 & 7 & 1.8\\ 
app10 & 4233  & 972K    & 11880&	138&	12.7& 	11876	&111	&11.1 &	0.20& 8 & 2.18 \\ 
\hline
\end{tabular}
}
\label{table:industry}
\vspace*{-2ex}
\end{table*}

\section{Related Work} \label{sec:RelatedWork}

 Context sensitivity on inter-procedural analysis has been an important research topic for many years. In this section, we briefly review some of the past research efforts on this. Inter-procedural analysis can be classified into top-down analysis and compositional analysis.
 
\textbf{Top down analysis}. Inlining is a conventional way to perform inter-procedural analysis~\cite{cousot1977abstract}. However, full inlining can hardly scale up. Thus, calling contexts are often limited. For instance, in \emph{k}-CFA~\cite{shivers1988control,grove1997call}, the length of a call string  is limited to be \emph{k}. Various representations for adopting limited contexts have been proposed, including \emph{k}-object~\cite{milanova2005parameterized,smaragdakis2011pick}, \emph{k}-type~\cite{smaragdakis2011pick}, and \emph{k}-this~\cite{lundberg2009fast}. A popular way for implementing such an analysis is to use BDDs~\cite{whaley2004cloning,berndl2003points} and Datalog~\cite{bravenboer2009strictly}. 
Some work models inter-procedural analysis as a reachability problem~\cite{reps1995precise,sagiv1996precise,zhang2017context}, which can be solved
with a pushdown system~\cite{reps2005weighted}. Their work has been applied to various non-trivial analyses like pointer analysis~\cite{spath2019context,zhang2013fast,xu2009scaling,chatterjee2017optimal} and taint analysis~\cite{arzt2014flowdroid} with field sensitivity.

Given the large size of current industrial software,  top-down inter-procedural analysis can hardly scale up  with high context sensitivity (\ie a relatively large \emph{k}). Many  optimizations have been considered, including parallization~\cite{albarghouthi2012parallelizing}, demand driven analysis~\cite{sridharan2005demand,arzt2014flowdroid,heintze2001demand}, selective context sensitivity~\cite{jeong2017data,li2018precision,li2020principled,lu2019precision}, incremental analysis~\cite{shang2012fast}. Another useful optimization~\cite{wang2017graspan,zuo2021chianina} utilizes disks to store the facts generated. 

\textbf{Compositional analysis}. Compositional analysis has been shown to be scalable in the industry~\cite{calcagno2009compositional,distefano2019scaling}, since it allows every procedure to be analyzed only once and facilitates
both incrementalization~\cite{calcagno2009compositional} and parallelization~\cite{shi2020pipelining}. A theoretical investigation has been made in \cite{cousot2001compositional}. Some recent advances fall mainly on designing new symbolic abstract domains to summarize pointer information~\cite{dillig2011precise,whaley1999compositional,chatterjee1999relevant,nystrom2004bottom} and shape properties~\cite{gulavani2011bottom,illous2020interprocedural}.  Symbolic numeric domains like Polyhedra \cite{cousot1978automatic} can also enable compositional analysis. Compositional analyses~\cite{nystrom2004bottom,shi2018pinpoint,illous2020interprocedural,zhang2014hybrid} usually cooperate with top-down analyses for various purposes 
such as call graph construction, property checking or abstract state refinement.
These works are all orthogonal to our work. The main difference is that, our work is about \emph{when} to summarize  statements, while all theses works are about \emph{how} to summarize  statements.

Our work can be seen as an approach on selective context sensitivity. This topic has been well explored for top-down analysis~\cite{jeong2017data,li2018precision,li2020principled,lu2019precision}. However, different from all existing these works, our approach the first attempt on compositional analysis. Another difference is that, the context sensitivity in our approach is more fine-grained (i.e., in the level of statements, rather than procedures).

\presec
\section{Conclusion} \label{sec:conclusion}
\postsec

We present a \toolname framework, which inlines both the critical statements and the summarization of non-critical statements at each callsite. It combines the advantages of top-down analysis (context sensitivity) and compositional analysis (scalability). Our work represents the first attempt to apply selective context sensitivity on compositional analysis. What is more, our approach can assign context sensitivity in the level of statements, rather than procedures. This fine-grained context sensitivity alleviates performance loss. In our evaluation, the pointer analysis instantiated from our framework can accomplish context sensitivity with little performance loss (65\% additional time overhead on Dacapo and 1\% on industrial applications).


\bibliographystyle{ACM-Reference-Format}
\bibliography{popl}

\end{document}